\documentclass[12pt,leqno]{article}
\usepackage{amsmath,amsthm,amssymb}

\def\init{\setcounter{equation}{0}}
\newtheorem{theorem}{Theorem}[section]
\newcommand{\R}{\mathbb{R}}

\newcommand{\e}{{\varepsilon}}

\newcommand{\rw}{\rightarrow}

\usepackage{tikz}
\usetikzlibrary{decorations.pathreplacing}
\usepackage{verbatim}

\usepackage{hyperref}

\title{Superradiance initiated inside the ergoregion}

\author{Gregory  Eskin, \\   Department of Mathematics, UCLA, \\ Los Angeles,
CA 90095-1555, USA. \\ eskin@math.ucla.edu
}

\begin{document}
\maketitle

\begin{abstract}
We consider  the stationary metrics that have both the black hole and the ergoregion.  The class of such metric contains,  in particular,
the Kerr metric.  We study the Cauchy problem  with highly oscillatory initial  data supported in a neighborhood  
inside the ergoregion with some 
initial energy $E_0$.   We prove that when the time variable  $x_0$  increases   this solution  splits into two  parts:  one  with the negative 
 energy  $-E_1$  ending at  the event horizon  in a finite time,  and  the second  part,  with the energy  $E_2=E_0+E_1>E_0$,
 escaping,  under some conditions, 
 to the infinity  when  $x_0\rw +\infty$.  Thus  we get  the superradiance  phenomenon.  In the case   of   
 the Kerr metric the superradiance phenomenon is ``short-lived",  since both the solutions with positive and negative energies 
 cross the outer event horizon 
  in a finite time (modulo $O(\frac{1}{k})$)  where  $k$  is a large parameter.  We show that these solutions end 
on the singularity ring in a finite time.  We study  also  the case  of naked  singularity.
\end{abstract}


\section{Introduction.}
\label{section 1}
\init
Consider the Lorentzian  metric in $\R^3\times \R$  of  the form
\begin{equation}                     												\label{eq:1.1}
dx_0^2-\sum_{k=1}^3dx_k^2- K(x)\Big(-dx_0+\sum_{k=1}^3 b_k(x)dx_k\big)^2,
\end{equation}
where $K(x)>0,\ \sum_j^3b_j^2(x)=1$.

The ergoregion for this metric  is the region where $K>1$.

Let  $(\rho,\varphi,z)$  be the cylindrical  coordinates  in $\R^3$,  i.e.  $x_1=\rho\cos\varphi,\ x_2=\rho\sin\varphi,\ x_3=z$,
let $x_0$  be  the time  variable.

Write the inverse  to the metric  tensor in the cylindrical coordinates.  Then the Hamiltonian $H(\rho,\varphi,z,\xi_0,\xi_\rho,\rho_\varphi,\xi_z)$
has the form
\begin{equation} 																	\label{eq:1.2}
H=\xi_0^2-\xi_\rho^2-\frac{1}{\rho^2}\xi_\varphi^2-\xi_z^2+K(-\xi_0 + \hat b\cdot \hat \xi)^2,
\end{equation}
where
$\hat\xi=(\xi_\rho,\frac{\xi_\varphi}{\rho},\xi_z), \hat b=(b_\rho,b_\varphi,b_z)$,
\begin{equation}															\label{eq:1:3}
|\hat b|^2=b_\rho^2+b_\varphi^2+b_z^2=1.
\end{equation}			
A particular  case of the metric (\ref{eq:1.1}) is the Kerr metric in Kerr-Schield coordinates  (cf. [14],  [15]):
\begin{equation}															\label{eq:1.4}
K=\frac{2mr^3}{r^4+a^2z^2},\ \ b_\rho=\frac{r\rho}{r^2+a^2},\ \ b_\varphi=\frac{a\rho}{r^2+a^2},\ \ b_z=\frac{z}{r},
\end{equation}
where $r(\rho,z)$  is defined by the relation
\begin{equation}																\label{eq:1.5}
\frac{\rho^2}{r^2+a^2}+\frac{z^2}{r^2}=1.
\end{equation}
Note that the ergosphere (outer ergosphere) for the Kerr metric  is 
$$
K=\frac{2mr^3}{r^4+a^2z^2}=1,
$$
and the event horizons (outer and inner event horizons)  are
\begin{equation}															\label{eq:1.6}
r_\pm=m\pm\sqrt{m^2- a^2},\ \ 0<a<m.
\end{equation}
Substituting  $r=r_\pm$  in (\ref{eq:1.5})  we see that the equation  of the event  horizon  is an ellipse in $(\rho,z)$ coordinates.

Note that  we do not consider  the full  Kerr  spacetime  but only  a part  restricted  to $\R^3\times \R$  (cf.  [6],  [15]).

Another  example  of the metric of the form  (\ref{eq:1.1})  is the Gordon's metric  (cf. [8], [9])  describing the propagation  of light  
in a moving dielectric  medium:
\begin{equation}																\label{eq:1.7}
dx_0^2-\sum_{j=1}^3dx_j^2 +(n^2-1)\Big(\sum_{j=0}^3v^{(j)}dx_j\Big)^2,
\end{equation}
where  $n(x)$  is  the reflection index, $v^{(0)}=\frac{1}{(1-\frac{|w|^2}{c^2})^{\frac{1}{2}}},
 v^{(j)}=\frac{w_j}{(1-\frac{|w|^2}{c^2})^{\frac{1}{2}}},1\leq j\leq 3,\linebreak w(x)=(w_1,w_2,w_3)    $
is  the velocity  of the flow,  $c$  is the speed  of  light in the vacuum.

In this paper we shall pay a special attention  to the acoustic  metric describing  the acoustic waves in a moving fluid flow  (cf. [13]). 
 Assuming,  for the simplicity of notation,
that  the speed  of the sound and the density  are equal to 1,   we have     the following Hamiltonian  in the polar  coordinates  $(\rho,\varphi)$
\begin{equation}																\label{eq:1.8}
H=\Big(\xi_0+\frac{A}{\rho}\xi_\rho+\frac{B}{\rho^2}\xi_\varphi\big)^2-\xi_\rho^2-\frac{1}{\rho^2}\xi_\varphi^2,
\end{equation}
 where
 \begin{equation}                                                									\label{eq:1.9}
 v=\frac{A}{\rho}\hat\rho +\frac{B}{\rho}\hat \varphi
 \end{equation}
 is the velocity 
 of the flow,  $\hat\rho=(\cos\varphi,\sin\varphi),\ \hat\varphi=(-\sin\varphi,\cos\varphi),$
 i.e.  $\frac{A}{\rho}\hat\rho$  is the radial  component  of $v$
   and $\frac{B}{\rho}\hat \varphi$   is the angular  component of  $v$.
 
 This paper studies the phenomenon  of  superradiance for  the spacetimes  having a black hole and ergosphere.   The region
 between  the ergosphere  and the event horizon is called  the ergoregion.  We briefly  describe  the main  contributions to
 the superradiance phenomenon.
 
 On the level of     
particles this phenomenon was discovered by R. Penrose  [11].

He proposed the following thought experiment:

Suppose a particle with an energy $\e_0$  enters the ergoregion and somehow splits into two particles with energies $\e_1$  and $\e_2$.  
By the conservation of energy $\e_0=\e_1+\e_2$.   It may  happen  inside the ergoregion that 
$\e_2<0$.   The particle  with  the negative energy  enters  the black hole  at some time and the particle with positive  energy  travels to the
infinity.  Since  $\e_2<0$  we have  that  $\e_1=\e_0-\e_2>\e_0$,  i.e.  the particle  leaving  the ergoregion  
has an energy larger  than $\e_0$.

This phenomenon is called the extraction of energy from the black hole.

On the level of the waves,  i.e.  solutions of the wave equation,  Ya. Zeldovich [16]  and his
then  student A. Starobinsky [12]  considered
the radial component   of a single mode of the  wave equation after the separation of variables.
For this  radial component  they studied the  scattering on the real  line  and found that  the outgoing  energy  flux  may be larger  than  the incoming 
energy flux.  This is the superradiance on the level of modes.  Finster, Kamran,  Smoller and Yau  [7] 
gave a far reaching 
generalization of the result of [12]  to the level of wave packets.  They constructed a solution of the Cauchy problem for the wave equation
with the initial  data far from the ergosphere  such that  the restriction of the energy to the complement  of a neighborhood  of the black 
hole has the limit  when $x_0\rw +\infty$ larger than the energy of the initial data.  Recently,  Dafermas, Rodnianski  and Shlyapentoch [2] 
undertook  a comprehensive  study  of the time-dependent scattering theory in the case of the Kerr metric.  As a by-product of this theory
they obtained the superradiance phenomenon.  
A comprehensive survey  of the results on the superradiance  is given  in a recent  book  by Brito,  Cardoso and  Pani  [1].

Our approach  to the superradiance phenomenon  is in the spirit of the Penrose approach.

In \S 2 we consider the Cauchy problem  for the wave equation with highly oscillatory initial  data containing  a large parameter $k$  and
supported in a small neighborhood $U_0$  of some point $P_0$ in the ergoregion.   The highly oscillatory initial data allow 
to construct a geometric optics type solution $u(x_0,x)$ of the Cauchy problem  concentrated in a neighborhood of two null-geodesics
$\gamma_+$  and  $\gamma_-$
starting  at the point $P_0$. In \S 3  we get that the energy  integral  $E_{x_0}(u)$  is equal  to $E_{x_0}(u_0^+)+E_{x_0}(u_0^-)$  modulo
lower order  terms in $k$,  where  $u_0^+$  and  $u_0^-$ are supported  in a neighborhood  of $\gamma_+$  and $\gamma_-$,
respectively.  

When the neighborhood $U_0$  is contained in the ergoregion we can arrange  the initial data 
in $U_0$  such that $E_{x_0}(u_0^-)<0$  and  therefore  $E(u_0^+)>E_0(u)$  for large  $k$,  i.e.  the superradiance  takes place.
Note  that the results of  \S 2  and  \S 3  
can be applied  to a general  wave equation  (\ref{eq:2.7}) corresponding to a general  Lorentzian metric having
an ergorigion.
  In \S 4
we study the acoustic metric  with the angular  velocity  much larger  than the radial velocity (see (\ref{eq:1.9})).  We prove that $u_0^+$
escapes to the infinity when $x_0\rw +\infty$   and  $u_0^-$  crosses the event  horizon  in a finite time.
In \S 5  we study the acoustic metric  when  the  angular component  of the velocity  flow is less than the radial  component,  
i.e.  $|B|<|A|$.  In this  case  both  $u_0^+$  and   $u_0^-$  cross the event horizon   in a finite time.

In \S6 we consider  the case   of a white hole,  i.e.  $A>0$.   Note that the reversal  of the time   $x_0$  to  $-x_0$  transforms  the black  hole   into  a white hole.
Therefore,  if we consider  the Cauchy problem with the same  initial   data   on  $(-\infty,0]$  instead
  of  $[0,+\infty)$  we get  that  $u_0^+$  
tends to  infinity when  $x_0\rw -\infty$  and  $u_0^-$  approaches  the event  horizon  when   $x_0\rw -\infty$.

In \S 7  we consider the case  of the Kerr metric   assuming  that the initial  point  $P_0$  and  the two  null-geodesics
$\gamma_+$  and  $\gamma_-$  lie in the equatorial  plane.
This case  is  similar  to the acoustic  case when  $|B|<|A|$.  As  in \S 5  we get  that
the superradiance phenomenon is ``short-lived"  since 
  $u_0^+$  and  $u_0^-$  disappear inside
the black  hole after  a finite  time.

In \S 8  we study the behaviour of $u_0^+$  and  $u_0^-$  inside the event horizon  and we prove  that in the case
of the Kerr metric  both $u_0^+$  and  $u_0^-$  end on the singularity ring.

Finally,  in  \S 9  we  study the Kerr  metric  when $a^2=m^2$  (the extremal case)  and  when  $a^2>m^2$  (the case  of naked  singularity).  We  show that the behavior of  $u_0^+$  and  $u_0^-$  does not change significantly
even  when there is no black hole.

\section{Geometric  optics  type  solutions of the Cauchy problem}
\label{section 2}
\init
Consider  the system  of null-bicharacteristics for the Hamiltonian  (\ref{eq:1.2}):
\begin{align}														\label{eq:2.1}
&\frac{dx_0}{ds}=\frac{\partial H}{\partial \xi_0},\ x_0(0)=0,\  \frac{d\rho}{ds}=\frac{\partial H}{\partial \xi_\rho},\ \ \rho(0)=\rho',
\\
\nonumber
&\frac{d\varphi}{ds}=\frac{\partial H}{\partial \xi_\varphi}, \  \varphi(0)=\varphi',
\  \frac{dz}{ds}=\frac{\partial H}{\partial\xi_z},\ z(0)=z',
\  \frac{d\xi_\rho}{ds}=-\frac{\partial H}{\partial \rho},\  \xi_\rho(0)=\eta_\rho,
\\
\nonumber
&\frac{d\xi_\varphi}{ds}=-\frac{\partial H}{\partial \varphi},  \xi_\varphi(0)=\eta_\varphi,
\frac{d\xi_z}{ds}=-\frac{\partial H}{\partial z},\xi_z(0)=\eta_z,
 \frac{d\xi_0}{ds}=-\frac{\partial H}{\partial x_0}=0,\xi_0(0)=\eta_0.
\end{align}
Note that $\frac{\partial H}{\partial x_0}=0$   since the metric is stationary.

The bicharacteristic (\ref{eq:2.1})  is called  the null-bicharacteristic if for any $s\in \R$
\begin{align}															\label{eq:2.2}
& H(\rho(s),\varphi(s),z(s),\xi_0(s),\xi_\rho(s),\xi_\varphi(s),\xi_z(s))
\\
\nonumber
&=\xi_0^2-\hat\xi\cdot\hat\xi +K(-\xi_0+\hat b\cdot\hat\xi)^2=0
\end{align}
along the bicharacteristic.  Since the metric is stationary,  to have (\ref{eq:2.2}) for all s  it is enough to have
\begin{equation}                                 									\label{eq:2.3}
\eta_0^2-\hat\eta\cdot\hat\eta+K(\rho',\varphi',z')(-\eta_0+\hat b(\rho',\varphi',z')\cdot \hat\eta)^2=0,
\end{equation}
where $\hat\eta=(\eta_\rho,\frac{\eta_\varphi}{\rho'},\eta_z)$.

Equation (\ref{eq:2.2})  is a quadratic equation  in $\xi_0$:
\begin{equation}														\label{eq:2.4}
(1+K)\xi_0^2-2K(\hat b\cdot\hat\xi)\xi_0+K(\hat b\cdot\hat \xi)^2-\hat\xi\cdot\hat \xi=0.
\end{equation}
It has two  distinct  real  roots $\xi_0^\pm=\lambda^\pm(\rho,\varphi,z,\xi_\rho,\xi_\varphi,\xi_z)$,
where 
\begin{align}															\label{eq:2.5}
&\lambda^\pm=\frac{K\hat b\cdot \hat \xi\pm\sqrt {\Delta_1}}{1+K},\ \ \lambda^-<\lambda^+,
\\
																		\label{eq:2.6}
&\Delta_1=(1+K)\hat\xi\cdot\hat\xi-K(\hat b\cdot\hat\xi)^2.
\end{align}
We shall call the null-bicharacteristics corresponding  to  $\xi_0^\pm=\lambda^\pm$  the ``plus" (``minus")  null-bicharacteristics,
respectively.   The projection of null-bicharacteristic  on $(\rho,\varphi,z)$-space is called null-geodesic.

Consider  the wave equation corresponding   to the metric (\ref{eq:1.1}):
\begin{equation}														\label{eq:2.7}
\Box_gu(x_0,x) \overset{def}{=}
\sum_{j,k=0}^3\frac{1}{\sqrt{-g(x)}}\ \frac{\partial}{\partial x_j}\Big(\sqrt{-g}\ g^{jk}(x)\frac{\partial u(x_0,x)}{\partial x_k}\Big)=0,
\end{equation}
where $[g^{jk}]_{j,k=0}^3$  is the inverse  to the metric  tensor  in (\ref{eq:1.1}),  $g^{00}>0, g(x)=(\det[g^{jk}]_{j,k=0}^3)^{-1}$.
We shall construct  
geometric optic type  solutions  with initial value  supported in $U_0$   where $U_0$  is a small neighborhood  of some point
$P_0$  in the ergoregion.  Such geometric optics solutions are well-known  and can be found  in many references  (see,
for example,  the book  [3],  \S 64).

Let $\chi_0(\rho,\varphi,z)$  be a $C_0^\infty$ function  with the support  in  $U_0$ and  
  $y_0=(\rho_0,\varphi_0,z_0)$  be the coordinates of $P_0$.    
Let  $y'=(\rho',\varphi',z')$  be  any point  in $U_0$,  and
let  $\eta=(\eta_\rho,\eta_\varphi,\eta_z)$.

Denote by
\begin{equation} 															\label{eq:2.8}
x=x^\pm(x_0,y,\eta)
\end{equation}
the solution  of  (\ref{eq:2.1})   when  $\eta_0^\pm=\lambda^\pm(\rho',\varphi',z',\eta_\rho,\eta_\varphi,\eta_z)$   (cf.  (\ref{eq:2.5})),
i.e.  $x=x^+(x_0,y,t)$  is  a ``plus"  null-geodesics  and  $x=x^-(x_0,y,\eta)$  is  the ``minus"  null-geodesics.

Suppose that the Jacobian 
\begin{equation}															\label{eq:2.9}
\frac{D x^\pm}{Dy}(x_0,y,\eta)\neq 0
\end{equation}
for  $0\leq x_0\leq T^\pm,\ y\in U_0,\ \eta$  is fixed.  Then  
by  the inverse  function theorem  $y=y^\pm(x_0,x,\eta)$  and  $y^\pm(0,x,\eta)=x$.     The points
where $\frac{D x^\pm}{Dy}=0$  are called  focal points.   Assuming  (\ref{eq:2.9})  we have following theorem  (cf.  [3],  \S  64):
\begin{theorem} 															\label{theo:2.1}
Let  $k$  be a large  parameter.
There  exist solutions  $u^\pm(x_0,\rho,\varphi,z)$  of  (\ref{eq:2.7})  having the form
\begin{equation}             														\label{eq:2.10}
u^\pm=u_0^\pm+\frac{1}{k}v^\pm(x_0,\rho,\varphi,z,k),
\end{equation}
where 
\begin{equation}															\label{eq:2.11}
u_0^\pm=e^{ikS^\pm}a_0^\pm(x_0,\rho,\varphi,z,\eta_\rho,\eta_\varphi,\eta_z), 
\end{equation}
$v^+$  and its  derivatives  are uniformly bounded  on  $[0,T^\pm]$,  eikonals  $S^\pm$  satisfy  the equations
\begin{align}        															\label{eq:2.12}
&S_{x_0}^\pm-\lambda^\pm(\rho,\varphi,z,S_\rho,S_\varphi,S_z)=0,\ \ \ 0\leq x_0\leq T^\pm,
\\
\nonumber
&S^\pm\big|_{x_0=0}=\rho \eta_\rho+\varphi\eta_\varphi +z\eta_z.
\end{align}
and
\begin{equation}																	\label{eq:2.13}
a_0^\pm(\rho,\varphi,z,\eta_\rho,\eta_\varphi,\eta_z)=C^\pm(x_0,x,\eta)\chi_0(y^\pm(x_0,x,\eta)),
\end{equation}
where $C^\pm\neq 0$  and bounded,  $C^\pm(0,x,\eta)=1$.

Moreover,  the correction  term $v^\pm(x_0,\rho,\varphi,z,\eta_\rho,\eta_\varphi,\eta_z)$  can be chosen  such that 
$u^\pm=u_0^\pm +\frac{1}{k}v^\pm$  satisfies the following initial conditions: 
\begin{align}              															\label{eq:2.14}
&u^\pm\big|_{x_0=0}=\chi_0(\rho,\varphi,z)e^{ik(\rho\eta_\rho+\varphi\eta_\varphi+z\eta_z)},
\\
																				\label{eq:2.15}
&\frac{\partial u^\pm}{\partial x_0}\Big|_{x_0=0}=ik\lambda^\pm(\rho,\varphi,z,\eta_\rho,\eta_\varphi,\eta_z)
   \chi_0(\rho,\varphi,z)e^{ik(\rho\eta_\rho+\varphi\eta_\varphi+z\eta_z)}
\end{align}
\end{theorem}

It follows from  (\ref{eq:2.13})  that the support of $a_0^\pm$  in  $(x_0,\rho,\varphi,z)$ space is contained  in the union 
$W^\pm$  of all ``plus" (``minus") null-geodesics starting  on the support of $\chi_0(\rho,\varphi,z)$.
Therefore  $\mbox{supp}\,u_0^\pm\subset W^\pm$  and  $\mbox{supp}\,u^\pm$  is contained in $W^\pm$
modulo $O(\frac{1}{k})$.  
\\
\\
{\bf Remark 2.1.}
\\
We will need to  consider  also  the case  when  either  
$\frac{Dx^+}{Dy}(x_0,y_0,\eta)$  or $\frac{Dx^-}{Dy}(x_0,y_0,\eta)$
vanish  at a finite  number  of points  $x_0^{+,k},1\leq k\leq r_+$,  $x_0^{-,k}, 1\leq k\leq  r_-$,  i.e.  when  the focal  points  are 
present.  There  is a generalization  of geometric  optics  construction  (\ref{eq:2.11})  belonging  to Maslov  (cf  [10],  see also
[3],  \S 66),
that allows  to construct  solutions $u^\pm$  depending  on a large  parameter  $k$,   solving  initial  value  problem  (\ref{eq:2.14}),
(\ref{eq:2.15})  for  the equation  (\ref{eq:2.7})  on arbitrary  large  interval  $[0,T]$.
In the neighborhood  of the focal  point  $u^\pm=u_0^\pm+\frac{1}{k}v^\pm$,  where  $u_0^\pm$  has  a more complicate
form than  (\ref{eq:2.11})  (see  [10]  and [3], \S66 for details).
However outside  of the neighborhoods  of focal  points   $u_0^\pm$  can again  be represented  in a form  (\ref{eq:2.11}).  
It follows  from the construction  that  $\mbox{supp}\, u^+\subset  W^+$  modulo  terms of order  $O\big(\frac{1}{k}\big)$
and   $\mbox{supp}\, u^-\subset  W^-$  modulo   $O\big(\frac{1}{k}\big)$
as in the case  of the absence  of focal points.

Let
\begin{equation}																	\label{eq:2.16}
u=u^++u^-.
\end{equation}
Then  $u$  is an exact solution  of (\ref{eq:2.7})  satisfying  the initial conditions
\begin{equation}                    														\label{eq:2.17}
u\big|_{x_0=0}=u^+\big|_{x_0=0}+u^-\big|_{x_0=0}=2\chi_0e^{ik(\rho\eta_\rho+\varphi\eta_\varphi+z\eta_z)},
\end{equation}
\begin{equation}                    														\label{eq:2.18}
\frac{\partial u}{\partial x_0}\Big|_{x_0=0}=ik(\lambda^++\lambda^-)\chi_0e^{ik(\rho\eta_\rho+\varphi\eta_\varphi+z\eta_z)}.
\end{equation}

\section{Conservation  of  energy  integrals}
\label{section 3}
\init

If $u(x_0,x)$ is the solution  of (\ref{eq:2.7})  with some initial  conditions and $u(x_0,x)$ has a compact support  in $x$  then
\begin{equation}																	\label{eq:3.1}
 E_{x_0}(u)=\frac{1}{2}\int\limits_{\R^3}\Big(g^{00}\Big|\frac{\partial u(x_0,x)}{\partial x_0}\Big|^2-
 \sum_{j,k=1}^3g^{jk}(x)\frac{\partial u}{\partial x_j}\overline{\frac{\partial u}{\partial x_k}}\Big)\sqrt g dx
 \end{equation}
is independent of $x_0$.

To show this take the derivative  of the right  hand side of (\ref{eq:3.1})  in $x_0$.  Using that  $u(x_0,x)$  satisfies  (\ref{eq:2.7})  and 
integrating by parts  we get that $\frac{d}{dx_0}E_{x_0}(u)\equiv 0$,  i.e.  $E_{x_0}(u)=E_0(u)$.

In particular,  $E_{x_0}(u^\pm)=E_0(u^\pm)$.  Using the initial conditions  (\ref{eq:2.14}), (\ref{eq:2.15}) we get
\begin{equation}																	\label{eq:3.2}
E_0(u^\pm)=\frac{k^2}{2}\int\limits_{\R^3}\Big(g^{00}\Big(\frac{\partial S^\pm}{\partial x_0}\Big)^2
-\sum_{j,k=1}^3 g^{jk}\frac{\partial S^\pm}{\partial x_j}\frac{\partial S^\pm}{\partial x_k}\Big) \chi_0^2 \sqrt g dx.
\end{equation}
Note  that when  $x_0=0\ \ S_{x_j}^\pm=\eta_j,1\leq j\leq 3,\ S_{x_0}^\pm=\lambda^\pm(\rho,\varphi,z,\eta_\rho,\eta_\varphi,\eta_z)$.
\begin{theorem}																			\label{theo:3.1}
Let the metric  $\sum_{j,k=0}^3g_{jk}(x)dx_jdx_k$  has a nonempty  ergoregion  $\Omega$,  i.e.   the domain
where   $g_{00}(x)<0$. 
 Let  $P_0\in \Omega,\ y_0=(\rho_0,\varphi_0,z_0)$  be the coordinates  of  $P_0$,  and  $U_0$  is a small  neighborhood   
 of $P_0$  containing  in  $\Omega$.  One always  can find  $\eta=(\eta_\rho,\eta_\varphi,\eta_z)$  such  that
 $$
 \lambda^+(y_0,\eta)>\lambda^-(y_0,\eta)>0.
 $$
 Then,  assuming that  $U_0$  is small,  we have
 $$
 E_{x_0}(u^-)<0,\ \ \ E_{x_0}(u^+)>0,
 $$
 and 
 $$
 E_{x_0}(u^+)=E_{x_0}(u)-E_{x_0}(u^-)>E_{x_0}(u),
 $$
 where  $u=u^++u^-$.   Thus  the superradiance  takes place.  
\end{theorem}

{\bf Proof:}  It follows  from  (\ref{eq:2.12})  that
\begin{equation}																	\label{eq:3.3}
g^{00}(S_{x_0}^\pm)^2+2\sum_{j=1}^3g^{0j}S_{x_j}^\pm S_{x_0}^\pm+\sum_{j,k=1}^3g^{jk}S_{x_j}^\pm S_{x_k}^\pm=0,
\end{equation}
since  $H=(\xi_0-\lambda^+)(\xi_0-\lambda^-)$.

Therefore
\begin{equation}																	\label{eq:3.4}
-\sum_{j,k=1}^3g^{jk}S_{x_j}^\pm S_{x_k}^\pm=g^{00}(S_{x_0}^\pm)^2+2\sum_{j=1}^3g^{0j}S_{x_j}^\pm S_{x_0}^\pm.
\end{equation}
Substituting (\ref{eq:3.4})  into (\ref{eq:3.2})  we get
\begin{equation}																	\label{eq:3.5}
E_0(u^\pm)=k^2\int\limits_{\R^3}S_{x_0}^\pm\Big(g^{00}S_{x_0}^\pm+
\sum_{j=1}^3 g^{0j}\frac{\partial S^\pm}{\partial x_j}\Big) \chi_0^2 \sqrt g dx.
\end{equation}
It follows from (\ref{eq:3.3}),  analogously to (\ref{eq:2.5}),  (\ref{eq:2.6}),   that
\begin{equation}																	\label{eq:3.6}
g^{00} S_{x_0}^\pm +\sum_{j=1}^3 g^{0j}S_{x_j}^\pm=\pm\sqrt{\Delta_1},
\end{equation}
where 
\begin{equation}																	\label{eq:3.7}
\Delta_1=\Big(\sum_{j=1}^3g^{0j}S_{x_j}^\pm\Big)^2-g^{00} \sum_{j,k=1}^3g^{jk}S_{x_j}^\pm S_{x_k}^\pm.
\end{equation}
Note that  in the case of metric (\ref{eq:1.1})
 $\Delta_1$  has the form (\ref{eq:2.6})  and $g^{00}=1+K$.
 
 Therefore considering,  for definiteness,  the case  of metric (\ref{eq:1.1})  we have 
 \begin{equation}																     \label{eq:3.8}
 E_0(u^+)=k^2\int\limits_{\R^3}S_{x_0}^+\sqrt{\Delta_1}\chi_0^2\sqrt g dx
 =k^2\int\limits_{\R^3}\lambda^+\sqrt{\Delta_1}\chi_0^2\rho d\rho d\varphi dz,
 \end{equation}
\begin{equation}																	\label{eq:3.9}
E_0(u^-)=-k^2\int\limits_{\R^3}S_{x_0}^-\sqrt{\Delta_1}\chi_0^2\sqrt g dx
=-k^2\int\limits_{\R^3}\lambda^-\sqrt{\Delta_1}\chi_0^2\rho d\rho d\varphi dz,
\end{equation}
where
$x_0=0$  and  $S_{x_0}^\pm,\Delta_1^\pm$  have the form (\ref{eq:2.5}), (\ref{eq:2.6}).
Note that $E_{x_0}(u^\pm)=E_0(u^\pm)$.

Now  compute  $E_0(u^++u^-)$.

It follows  from  (\ref{eq:2.17}),  (\ref{eq:2.18})  and (\ref{eq:3.1})   that (cf. (\ref{eq:2.4}))
\begin{equation}																	\label{eq:3.10}
E_0(u^++u^-)=\frac{k^2}{2}\int\limits_{\R^3}\Big((1+K)(\lambda^++\lambda^-)^2\chi_0^2-
(K(\hat b\cdot\hat\eta)^2-\hat\eta\cdot\hat\eta)4\chi_0^2\Big)\rho d\rho d\varphi dz.
\end{equation}
Since
\begin{equation}																	\label{eq:3.11}
\lambda^+ +\lambda^-=\frac{2K(\hat b\cdot\hat\eta)}{1+K},
\end{equation}
we have
\begin{align}																		\label{eq:3.12}
E_0(u^++u^-)=&\frac{k^2}{2}\int\limits_{\R^3}\Big[(1+K)\frac{4K^2(\hat b\cdot\hat\eta)^2}{(1+K)^2}
-4(K(\hat b\cdot\hat\eta)^2- \hat\eta\cdot\hat \eta)\Big]\chi_0^2\rho d\rho d\varphi dz
\\
\nonumber
=&2k^2\int\limits_{\R^3}\Big(\frac{K^2(\hat b\cdot\hat\eta)^2}{1+K}
-K(\hat b\cdot\hat\eta)^2+\hat\eta\cdot\hat \eta\Big)\chi_0^2\rho d\rho d\varphi dz
\\
\nonumber
=&2k^2\int\limits_{\R^3}\frac{(1+K)\hat \eta\cdot\hat\eta- K(\hat b\cdot\hat\eta)^2}{1+K}
\chi_0^2\rho d\rho d\varphi dz
\\
\nonumber
=&2k^2\int\limits_{\R^3}\frac{\Delta_1}{1+K}
\chi_0^2\rho d\rho d\varphi dz
\end{align}
From the other side,
\begin{equation}															\label{eq:3.13}
E_0(u^+)+E_0(u^-)=k^2\int\limits_{\R^3}(\lambda^+\sqrt{\Delta_1}-\lambda^-\sqrt{\Delta_1}) \chi_0^2\rho d\rho d\varphi dz.
\end{equation}
Since $\lambda^+-\lambda^-=\frac{2\sqrt{\Delta_1}}{1+K}$  we are getting  that
$$
E_0(u^++u^-)=E_0(u^+)+E_0(u^-),
$$
and
thus for any  $x_0$  
\begin{equation}															\label{eq:3.14}
E_{x_0}(u^++u^-)=E_{x_0}(u^+)+E_{x_0}(u^-).
\end{equation}
When  $y=(\rho',\varphi',z')$  is outside  of the ergoregion  then $K<1$  and
$$
K(\hat b\cdot\hat \eta)^2-\hat\eta\cdot\hat\eta<0
$$ 
for all $\hat\eta\neq 0$.  Therefore  $\lambda_+$  and $\lambda_-$  have opposite  signs (cf.  (\ref{eq:2.4})).

When  $y_0=(\rho_0,\varphi_0,z_0)$  belongs  to the  ergoregion  then $K>1$  and one can  choose 
 $\hat\eta_0=(\eta_\rho,\eta_\varphi,\eta_z)$  such  that
$$
K(\hat b\cdot \hat \eta_0)^2-\hat\eta_0\cdot\hat\eta_0>0.
$$
Thus  $\lambda^+(y_0,\hat \eta_0)$  and  $\lambda^-(y_0,\hat \eta_0)$  have  the same  sign.

Without loss of generality  we can  assume   that  $\lambda_+(y_0,\hat\eta_0)>\lambda^-(y_0,\hat\eta_0)>0$.
Otherwise  we replace  $\hat\eta_0$  by  $-\hat\eta_0$  and use  that  $\lambda^+(y_0,\hat \eta_0)= -\lambda^-(y_0,-\hat \eta_0)$.

   Then  $E_{x_0}(u^-)<0$  when  $\mbox{supp}\,\chi_0$  contains 
in a small neighborhood of  $(\rho_0,\varphi_0,z_0)$  (see  (\ref{eq:3.9})).
\qed

Since  $E_{x_0}(u^++u^-)=E_{x_0}(u^+)+E_{x_0}(u^-)$  we have that
\begin{equation}															\label{eq:3.15}
E_{x_0}(u^+)=E_{x_0}(u^++u^-)-E_{x_0}(u^-)>E_{x_0}(u^++u^-)=E_0(u^++u^-).
\end{equation}

  Note that  $\mbox{supp}\, u_0^\pm\subset W^\pm$  (cf. (\ref{eq:2.13})),
where $W^+(W^-)$  is the union  of all null-geodesics  starting  at  $(y,\eta)$ where  $y\in \mbox{supp}\, \chi_0$.   Therefore,  for any 
 $x_0$  one can find $k$  so large  that 
\begin{equation}															\label{eq:3.16}
E_{x_0}(u_0^+)>E_{x_0}(u^++u^-)=E_0(u^++u^-).
\end{equation}
We took
into account 
(\ref{eq:3.15}) and that $\frac{E_{x_0}(u^+)}{E_{x_0}(u_0^+)}=1+O\big(\frac{1}{k}\big)$.  
Therefore  (\ref{eq:3.16})  show  that the superradiance  takes place.  Note that  the ``minus"   null-geodesics can not
leave  the ergoregion since $\xi_0^-=\lambda^-(y,\eta)>0$.  
It will be shown  that
when  $x_0>T_1\ \ $  $u_0^-$  is inside  the event horizon.  Therefore,
outside the event horizon $\frac{E_{x_0}(u^-)}{E_{x_0}(u_0^-)}=O\big(\frac{1}{k}\big)$    for  $x_0$  large.
\\
\\
{\bf Remark 3.1.  The case of real-valued solutions.}
Since
coefficients  of  equation  (\ref{eq:2.7})  are  real,  the real  (and  imaginary)  part  of complex-valued solution  also satisfies (\ref{eq:2.7}).
Let  $u^\pm=u_0^\mp +\frac{1}{k}v^\pm$  be the same  as  in  (\ref{eq:2.10}).   Denote  by  
$u_R^\pm=u_{0R}^\mp +\frac{1}{k}v_R^\pm$
the real part  of  (\ref{eq:2.10}).  Since $a_0^\pm$  is  real-valued  we have
\begin{equation}																\label{eq:3.17}
u_R^\pm=\frac{1}{2}(u^\pm+\overline{u^\pm})\equiv a_0^\pm\cos kS^\pm + O\Big(\frac{1}{k}\Big).
\end{equation}
Moreover,  (\ref{eq:2.14}),  (\ref{eq:2.15})  imply
\begin{align}              															\label{eq:3.18}
&u_R^\pm\big|_{x_0=0}=\chi_0(\rho,\varphi,z)\cos k (\rho\eta_\rho+\varphi\eta_\varphi+z\eta_z),
\\
\nonumber																				
&\frac{\partial u_R^\pm}{\partial x_0}\Big|_{x_0=0}=-k\lambda^\pm(\rho,\varphi,z,\eta_r,\eta_\varphi,\eta_z)
   \chi_0(\rho,\varphi,z)\sin k (\rho\eta_\rho+\varphi\eta_\varphi+z\eta_z).
\end{align}
Since  (\ref{eq:3.1})  holds   
for real-valued  solutions  too  and since  (\ref{eq:3.1})   is independent  of  $x_0$,  we  have,  as in  (\ref{eq:3.2}):  
\begin{multline}																	\label{eq:3.19}
E_0(u_R^\pm)
\\
=\frac{k^2}{2}\int\limits_{\R^3}\Big(g^{00}\Big(\frac{\partial S^\pm}{\partial x_0}\Big)^2
-\sum_{j,k=1}^3 g^{jk}\frac{\partial S^\pm}{\partial x_j}\frac{\partial S^\pm}{\partial x_k}\Big) \chi_0^2 
\sin^2 k (\rho\eta_\rho+\varphi\eta_\varphi+z\eta_z)
\rho d\rho d\varphi.
\end{multline}
Therefore,  as  in (\ref{eq:3.8}),  (\ref{eq:3.9}),  we get 
 \begin{align}																     \label{eq:3.20}
 &E_0(u_R^+)
 =k^2\int\limits_{\R^3}\lambda^+\sqrt{\Delta_1}
 \sin^2 k(\rho\eta_\rho+\varphi\eta_\varphi+z\eta_z)
 \chi_0^2\rho d\rho d\varphi dz >0,
\\
\label{eq:3.21}
&E_0(u_R^-)
=-k^2\int\limits_{\R^3}\lambda^-\sqrt{\Delta_1}
\sin^2 k(\rho\eta_\rho+\varphi\eta_\varphi+z\eta_z)
\chi_0^2\rho d\rho d\varphi dz<0.
\end{align}
Since $\sin^2 k(\rho\eta_\rho+\varphi\eta_\varphi+z\eta_z)=\frac{1}{2} -
\frac{1}{2}\cos 2k(\rho\eta_\rho+\varphi\eta_\varphi+z\eta_z)$   and since integrals  containing 
$ \cos 2 k(\rho\eta_\rho+\varphi\eta_\varphi+z\eta_z)$  are  decaying fast  when  $k\rw \infty$  we have,  as in  
(\ref{eq:3.14}),  (\ref{eq:3.15}),  (\ref{eq:3.16}),   that
\begin{equation}																			\label{eq:3.22}
E_{x_0}(u_{0R}^+)>E_{x_0}(u_R^++u_R^-)=E_0(u_R^++u_R^-)
\end{equation}
when $k$  is large.

Thus superradiance  holds  also  for real-valued  solutions.
Note that  
$$
E_{x_0}(u)=E_{x_0}(u_R)+E_{x_0}(u_I),
$$
where  $u=u_R+iu_I,, u_R=\Re u, u_I=\Im u$.
Also  $E_{x_0}(u_R)\approx \frac{1}{2}E_{x_0}(u),\  E_{x_0}(u_I)\approx \frac{1}{2} E_{x_0}(u)$  when  $k\rw \infty$.

\section{Superradiance in the case  of the acoustic metric}
\label{section 4}
\init

Consider in more details the case of the acoustic  metric  (\ref{eq:1.8})  (cf.  [13],  [4],  [5]).
  The equation  $H=0$  (cf. (\ref{eq:1.8}))  
has two roots
\begin{equation}																				\label{eq:4.1}
\xi_0^\pm=-\frac{A}{\rho}\xi_\rho - \frac{B}{\rho^2}\xi_\varphi \pm
 \sqrt{\Delta_1},
\end{equation}
where
\begin{equation}																				\label{eq:4.2}
\Delta_1=\xi_\rho^2 +\frac{1}{\rho^2}\xi_\varphi^2.
\end{equation}
The black hole is $\{\rho<|A|,\ A<0\}$,   and  the ergoregion  is $\{|A|<\rho<\sqrt{A^2+B^2}\}$.    We assume,  for definiteness,
$B>0$.  

We shall find  first  simple expressions  for the differential  equations  for  the "plus"  and  "minus"   null-geodesics.
Note that
\begin{align}
\nonumber
&H_{\xi_\rho}=2\Big(\xi_0+\frac{A}{\rho}\xi_\rho + \frac{B}{\rho^2}\xi_\varphi\Big)\frac{A}{\rho}-2\xi_\rho,
\\
\nonumber
&H_{\xi_0}=2\Big(\xi_0+\frac{A}{\rho}\xi_\rho + \frac{B}{\rho^2}\xi_\varphi\Big).
\end{align}
It follows  from  (\ref{eq:2.1})
that  in the acoustic case
\begin{equation}																				\label{eq:4.3}
\frac{d\rho^\pm}{d x_0}=\frac{H_{\xi_\rho}}{H_{\xi_0}}=\frac{A}{\rho}-\frac{\xi_\rho}{\xi_0^\pm+\frac{A}{\rho}\xi_\rho+\frac{B\xi_\varphi}{\rho^2}}
=\frac{(\frac{A^2}{\rho^2}-1)\xi_\rho +\frac{A}{\rho}(\xi_0^\pm+\frac{B\xi_\varphi}{\rho^2})}
{\xi_0^\pm+\frac{A}{\rho}\xi_\rho+\frac{B\xi_\varphi}{\rho^2}}.
\end{equation}
Since  $H=0$,  the following  quadratic  equation  for  $\xi_p$ holds:
\begin{equation}																				\label{eq:4.4}
\Big(\frac{A^2}{\rho^2}-1\Big)\xi_\rho^2+2\frac{A}{\rho}\Big(\xi_0^\pm+\frac{B\xi_\varphi}{\rho^2}\Big)\xi_\rho
+\Big(\xi_0^\pm+\frac{B\xi_\varphi}{\rho^2}\Big)^2-\frac{1}{\rho^2}\xi_\varphi^2=0.
\end{equation}
Therefore
\begin{equation}																				\label{eq:4.5}								
\xi_\rho=\frac{-\frac{A}{\rho}\big(\xi_0^\pm+\frac{B\xi_\varphi}{\rho^2}\big)\pm\sqrt{\Delta_2^\pm}}{\frac{A^2}{\rho^2}-1},
\end{equation}
where
\begin{multline}																				\label{eq:4.6}
\Delta_2^\pm=\frac{A^2}{\rho^2}\Big(\xi_0^\pm+\frac{B\xi_\varphi}{\rho^2}\Big)^2-\Big(\frac{A^2}{\rho^2}-1\Big)\Big(\xi_0^\pm +
\frac{B\xi_\varphi}{\rho^2}\Big)^2+\Big(\frac{A^2}{\rho^2}-1\Big)\frac{\xi_\varphi^2}{\rho^2}
\\
=\Big(\xi_0^\pm +\frac{B\xi_\varphi}{\rho^2}\Big)^2+\Big(\frac{A^2}{\rho^2}-1\Big)\frac{\xi_\varphi^2}{\rho^2}.
\end{multline}
Substituting (\ref{eq:4.5})  in the numerator  of (\ref{eq:4.3})  we get
\begin{equation}																			\label{eq:4.7}
\frac{d\rho^\pm}{dx_0}=\frac{\pm\sqrt{\Delta_2^\pm}}{\xi_0^\pm+\frac{A}{\rho}\xi_r+\frac{B\xi_\varphi}{\rho^2}}=
\frac{\pm\sqrt{\Delta_2^\pm}}{\pm\sqrt{\xi_\rho^2+\frac{1}{\rho^2}\xi_\varphi^2}}.
\end{equation}
Therefore
\begin{equation} 																			\label{eq:4.8}
\frac{d\rho^+}{dx_0}=\frac{\pm\sqrt{\Delta_2^\pm}}{\sqrt{\Delta_1}},\ \ \ \ \ \ \
 \frac{d\rho^-}{dx_0}=\frac{\pm\sqrt{\Delta_2^\pm}}{-\sqrt{\Delta_1}}
\end{equation}
In this section  we consider  the case when  the angular velocity  is much larger  than the radial  velocity,
i.e.  $B\gg |A|$.   The following theorem  holds:
\begin{theorem}																				\label{theo:4.1}
Let $2|A|<\rho_0<\sqrt{A^2+B^2}$.  Choose  $\eta_\rho=-\frac{2|A|}{\rho_0},\eta_\varphi=-\rho_0\sqrt{1-\frac{4A^2}{\rho_0^2}}$,
i.e.  $\eta_\rho^2+\frac{\eta_\varphi^2}{\rho_0^2}=\frac{4A^2}{\rho_0^2}+\big(1-\frac{4A^2}{\rho^2}\big)=1$.
Suppose
\begin{equation}																			\label{eq:4.9}
B>\Big(1+\frac{2A^2}{\rho_0^2}\Big)\frac{\rho_0}{\sqrt{1-\frac{4A^2}{\rho_0^2}}}.
\end{equation} 
Let  $u^\pm=u_0^\pm +\frac{1}{k}v^\pm$  be the  same  as in  (\ref{eq:2.11}),  (\ref{eq:2.12})  satisfying  the  initial conditions
(\ref{eq:2.14}),  (\ref{eq:2.15}).

Then the superradiance takes  place,  i.e.  
$E_{x_0}(u_0^+)> E_0(u^++u^-)$  for large  $k,\ \mbox{supp}\, u_0^+$  tends to infinity  when  $x_0\rw +\infty$,
  $\mbox{supp}\, u_0^-$  crosses  the event horizon  $\rho= |A|$  when  $x_0$ increases.
\end{theorem}
{\bf Proof:}
Find first the sign  of $\frac{d\rho^{\pm}}{dx_0}$  at  the initial point  
$(\rho_0,\varphi_0),\ \xi_\rho=\eta_\rho,\ \xi_\varphi=\eta_\varphi$  when  $x_0=0$.
Since $\rho_0>2|A|,\ \eta_\rho=-\frac{2|A|}{\rho_0}, \ \eta_\varphi=-\rho_0\sqrt{1-\frac{4A^2}{\rho_0^2}}$,
 we have
\begin{multline}																					\label{eq:4.10}
\xi_0^-\big|_{x_0=0}=-\frac{A}{\rho_0}\eta_\rho-\frac{B}{\rho_0^2}\eta_\varphi-\sqrt{\Delta_1}=
-\frac{2A^2}{\rho_0^2}+\frac{B}{\rho_0}\sqrt{1-\frac{4A^2}{\rho_0^2}}-1.
\end{multline}
If (\ref{eq:4.9}) holds 
then $\xi_0^->0$   and therefore the superradiance  will take place. Also
\begin{equation}																				\label{eq:4.11}
\frac{d\rho^+}{dx_0}\Big|_{x_0=0}=\frac{A}{\rho_0}-\frac{\eta_\rho}
{\sqrt{\eta_\rho^2+\frac{\eta_\varphi^2}{\rho_0^2} }}
=-\frac{|A|}{\rho_0}+2\frac{|A|}{\rho_0}>0,
\end{equation}
since  
$\sqrt{\eta_\rho^2+\frac{\eta_\varphi^2}{\rho_0^2}}=1,\ \eta_\rho=-\frac{2|A|}{\rho_0}$.  Therefore
\begin{equation}																				\label{eq:4.12}
\frac{d\rho^+}{dx_0}=\frac{\sqrt{\Delta_2^+}}{\sqrt{\Delta_1}}>0\ \  \mbox{for  all}\ \  x_0\geq 0,
\end{equation}
if we can prove that $\Delta_2^+>0$  for all $\rho\geq \rho_0$,   (zeros  of  $\Delta_2^\pm$   are called the turning points).

Rewrite  equation  $\Delta_2^\pm=0$  (cf.  (\ref{eq:4.6}))  in the form
\begin{equation}																				\label{eq:4.13}
\frac{(A^2+B^2)\xi_\varphi^2}{\rho^4}+\frac{2\xi_0^\pm B\xi_\varphi-\xi_\varphi^2}{\rho^2}+(\xi_0^\pm)^2=0,
\end{equation}
where
\begin{equation}																				\label{eq:4.14}
\xi_0^\pm\big|_{x_0=0}=-\frac{2A^2}{\rho_0^2}- \frac{B\eta_\varphi}{\rho_0^2}\pm 1.
\end{equation}
Solving  the quadratic  equation  in $\frac{1}{\rho^2}$  we get
\begin{equation}																				\label{eq:4.15}
\frac{1}{\rho^2}=\frac{-2\xi_0^\pm B \xi_\varphi +\xi_\varphi^2\pm\sqrt\delta}{2(A^2+B^2)\xi_\varphi^2}
\end{equation}
where 
\begin{align}																				\label{eq:4.16}
\delta&=(2B\xi_\varphi\xi_0^\pm-\xi_\varphi^2)^2-4(A^2+B^2)\xi_\varphi^2(\xi_0^\pm)^2
\\
\nonumber
&=-4B\xi_\varphi^3\xi_0^\pm +\xi_\varphi^4-4A^2\xi_\varphi^2(\xi_0^\pm)^2.
\end{align}
Since $B$  is large  (cf.  (\ref{eq:4.10}),  $\xi_\varphi=\eta_\varphi<0$  and  $\xi_0^\pm$  has  the form  (\ref{eq:4.14}),  we 
have
$$
\delta=4\xi_\varphi^2|B\xi_\varphi|\xi_0^\pm-4A^2\xi_\varphi^2(\xi_0^\pm)^2
+\xi_\varphi^4=4\xi_\varphi^2\frac{(B\xi_\varphi)^2}{\rho_0^2}
-4\xi_\varphi^2\frac{A^2(B\xi_\varphi)^2}{\rho_0^4}+\cdots,
$$
where $\cdots$  means  nonsignificant  terms.  Therefore
\begin{equation}																			\label{eq:4.17}
\sqrt\delta=2\frac{B\xi_\varphi^2}{\rho_0}\Big(1-\frac{A^2}{\rho_0^2}\Big)^{\frac{1}{2}}+\cdots
\end{equation}
Substituting in (\ref{eq:4.15})  and taking into the account  that $B\gg |A|$  we obtain
\begin{align}																				\label{eq:4.18}
\frac{1}{\rho^2}=&\frac{2|B\xi_\varphi|\big(\frac{B|\xi_\varphi|}{\rho_0^2}-\frac{2A^2}{\rho_0^2}\pm 1\big) \pm 
\frac{2B\xi_\varphi^2}{\rho_0}\big(1-\frac{A^2}{\rho_0^2}\big)^{\frac{1}{2}}}{2 B^2\xi_\varphi^2}+\cdots
\\
\nonumber
=
&\frac{1}{\rho_0^2}
+\frac{1}{B|\xi_\varphi|}\Big(\pm1 - \frac{2A^2}{\rho_0^2}\Big)        \pm \frac{\big(1-\frac{A^2}{\rho_0^2}\big)^{\frac{1}{2}}}{B\rho_0}+\cdots
\end{align}
Note  that  $|\xi_\varphi|=\rho_0\big(1-\frac{4A^2}{\rho_0^2}\big)^{\frac{1}{2}}$.  

It follows  from (\ref{eq:4.18})  that  $\Delta_2^+=0$  has  turning points  $\rho_2^+<\rho_1^+<\rho_0$,
where   $\rho_j^+$  solves
\begin{equation} 																	\label{eq:4.19}
\frac{1}{(\rho_j^+)^2}=\frac{1}{\rho_0^2}
+\frac{1}{B|\xi_\varphi|}\Big(1 - \frac{2A^2}{\rho_0^2}\Big)        \pm \frac{\big(1-\frac{A^2}{\rho_0^2}\big)^{\frac{1}{2}}}{B\rho_0}
+\dots,\ \ \ 
j=1,2,
\end{equation}
$\rho_1^+$  corresponding  to the minus  sign  in  (\ref{eq:4.19})  and  $\rho_2^+$  to  the plus sign.
Analogously,   $\Delta_2^-=0$  has turning points  $\rho_0<\rho_1^-<\rho_2^-$,  where $\rho_j^-$  solves
\begin{equation} 																	\label{eq:4.20}
\frac{1}{(\rho_j^-)^2}=\frac{1}{\rho_0^2}
+\frac{1}{B|\xi_\varphi|}\Big(-1 - \frac{2A^2}{\rho_0^2}\Big)        \pm \frac{\big(1-\frac{A^2}{\rho_0^2}\big)^{\frac{1}{2}}}{B\rho_0}
+\dots,\ \ \
j=1,2.
\end{equation}

Therefore $\Delta_2^+>0$  for all  $\rho>\rho_1^+$  and hence  $\frac{d\rho^+}{dx_0}>0$  for  all  $x_0>0$.    Thus  
$\rho^+(x_0)\rw +\infty$  when  $x_0\rw +\infty$.

Note that  $\frac{d\rho^-}{dx_0}\big|_{x_0=0}=-\frac{|A|}{\rho_0}
+\frac{\eta_\rho}{\sqrt{\eta_\rho^2+\frac{\eta_\varphi^2}{\rho_0^2}}}
=-\frac{3|A|}{\rho_0}<0.$

Thus  $\Delta_2^->0$  for $\rho<\rho_1^-$.  We have  that
\begin{equation}																				\label{eq:4.21} 
\frac{d\rho^-}{dx_0}=-\frac{\sqrt{\Delta_2^-}}{\sqrt{\Delta_1}}<0\ \  \mbox{for}\ \  x_0>0,
\end{equation}
 and $\rho^-(x_0)$  crosses  the event horizon 
when  $x_0$  increases.
Let
\begin{equation}																				\label{eq:4.22}    
u^\pm=e^{ikS^\pm}a_0^\pm+\frac{1}{k}v^\pm(x_0,\rho,\varphi,k)
\end{equation}
be the solutions  (\ref{eq:2.10}),  (\ref{eq:2.11})  in the case   of the acoustic  metric  and let  $E_{x_0}(u^\pm)$  be the corresponding
energy integrals  (cf.  (\ref{eq:3.8}),  (\ref{eq:3.9})).

When  the condition (\ref{eq:4.9})  holds then  
$\rho^+(x_0)\rw+\infty$  if $x_0\rw +\infty$.  Hence  $\mbox{supp}\, u_0^+$  tends  to the infinity  
when  $x_0\rw +\infty$.
Also  $\mbox{supp}\, u_0^-$  enters  the  black   hole  $\rho=|A|$  when  $x_0$  increases.
As in  (\ref{eq:3.16})  we have  $E_{x_0}(u_0^+)>E_0(u^++u^-)$  for large  $k$,  i.e.  the superradiance takes place.
\qed
\\
\\
{\bf Remark 4.1.  The behavior  of  ${\bf u_0^+}$  and  ${\bf u_0^-}$   on the time interval (${\bf -\infty,0}$).}
\\
Consider  the behavior of $\rho^+(x_0)$   and  $\rho^-(x_0)$  on the time interval  $(-\infty,0]$.  Since $\rho_1^+<\rho_0$  is
the turning  point  where  $\Delta_2^+=0$,  we  get that  $\rho^+(x_0)$  decreases  when  $x_0$  decreases  until it reaches  
$\rho_1^+$ 
at some  time  $x_0^{(1)}<0$.  
 Note  that the null-geodesics  $\gamma^+$  has a singularity  at  $\rho^+(x_0^{(1)})=\rho_1^+$  since  $\sqrt{\Delta_2^+}=
 C\sqrt{\rho^+(x_0)-\rho^+_1}$  for  $x_0>x_0^{(1)}$.  Passing the point  $x_0^{(1)}$    we  have  $ \frac{d\rho^+(x_0)}{dx_0}  =
 -\frac{\sqrt{\Delta_2^+}}{\sqrt{\Delta_1^+}}<0$  for  $x_0<x_0^{(1)}$  (cf. [4]),  i.e.   $\rho^+(x_0)$   turns  and  starts   increasing  when  
 $x_0<x_0^{(1)}$  decreasing.
 Since  $\Delta_2^+>0$  on  $(\rho_1^+,+\infty)$   we get  that  $\rho^+(x_0)\rw +\infty$  when  $x_0\rw -\infty,  x_0<x_1^{(0)}$.
 Similarly,  for  $\rho^-(x_0)$  the  point  $\rho_1^->\rho_0$    is also the 
 turning point  and
$\rho^-(x_0)$  increases when $x_0$  decreases until it reaches $\rho_1^-$  at some  time $x_0^{(2)}<0$.
Then  $\rho^-(x_0)$  turns and starts  to decrease when  $x_0$  decreases.  When  $x_0\rw -\infty\ \ \ \rho^-(x_0)$  tends  to the event 
 horizon $\rho=|A|$  and  $\rho^-(x_0)$ spirals  around  the event horizon  as $x_0\rw -\infty$   (cf. [ 4]).  

Turning points are  singularities (focal)  points   for 
null-geodesics.  When  the null-geodesics is passing through the focal  point  one needs the Maslov  extension  
 of the  geometric  optics  construction
as stated in Remark 2.1.   Note that  the turning  points  $\rho_1^+$  and  $\rho_1^-$  are  the simplest   type 
of  singularities called the  fold singularities.  A detailed geometric optics construction  for the case  of such  singularities  is presented  in 
[3],  Example 66.1.

Since  $\rho^+(x_0)\rw \infty$  when  $x_0\rw -\infty$  we  have that  $\mbox{supp}\, u_0^+$  tends to $\infty$  when
$x_0\rw -\infty$.  Also  $\mbox{supp}\,u_0^-$  tends  to the event  horizon  when  $x_0\rw -\infty$  since  $\rho^-(x_0)$  
approaches $\rho=|A|$  when $x_0\rw -\infty$.
\\
\\
{\bf Remark 4.2.  The  case  of acoustic  metric with  ${\bf A=0}$   (the case of naked singulirity)}.
\\
Consider  the acoustic  metric  (\ref{eq:1.8})   when  $A=0$.   We assume   as above  $B>0,\eta_\varphi<0$.
The equation  of the  ergosphere  is  $\rho=B$  and  there  is no event  horizon,  i.e.  $\rho=0$  is a ``naked"  singularity. 

We choose the initial point  $\rho_0$  such that 
 $\xi_0^-=-\frac{B\eta_\varphi}{\rho_0^2}-\sqrt{\eta_\rho^2+\frac{\eta_\varphi^2}{\rho_0^2}}>0$.  Thus
\begin{equation} 																\label{eq:4.23}            
\frac{(B^2-\rho_0^2)\eta_\varphi^2}{\rho_0^4}>\eta_\rho^2.
\end{equation}
It follows from  (\ref{eq:4.23})  that $\rho_0< B$,  i.e  $\rho_0$  
belongs  to the ergoregion.

We choose $\eta_\rho<0$  small and  $\eta_\rho$  satisfies  (\ref{eq:4.21})  and  choose  $\eta_\varphi$   such that
\begin{equation}				 												\label{eq:4.24}              
\frac{\eta_\varphi^2}{\rho_0^2}+\eta_\rho^2=1.
\end{equation}
Then 
$$
\xi_0^\pm=\frac{B|\eta_\varphi|}{\rho_0^2}\pm 1.
$$

We shall show  that  the  superradiance  takes  place  when  $A=0$  despite  
the absence  of the black hole.

\begin{theorem}																	\label{eq:4.2}
Suppose  $A=0,B>0,\rho_0,\eta_\varphi,\eta_\rho$  are    such that  $\rho_0<B,\eta_\rho<0,\eta_\varphi<0$
  and  (\ref{eq:4.23}),
(\ref{eq:4.24}) hold.
Let  $u_0^\pm=u_0^\pm+\frac{1}{k}v^\pm$  be the same  as in  Theorem  \ref{theo:4.1}  with  $\rho_0,\eta_\rho,\eta_\varphi$  
chosen as above  and
$\varphi_0$  is arbitrary.
Then  $E_0(u^-)<0$  since  $\xi_0^-=-\frac{B\eta_\varphi}{\rho_0^2} -  \sqrt{\eta_\rho^2+\frac{\eta_\varphi^2}{\rho_0^2}}>0$.
Thus  $E_{x_0}(u_0^+)>E_0(u^++u^-)$  for large  $k$, \linebreak i.e.  the superradiance holds.  Moreover,  $u_0^+$  escapes  to  the infinity
when  $x_0\rw +\infty$  and  $u_0^-$  approach  the  singularity  $\rho=0$  when  $x_0$  increases.
\end{theorem}
{\bf Proof:}
The equations (\ref{eq:4.3})  have the form
\begin{equation}																\label{eq:4.25}               
\frac{d\rho^\pm(x_0)}{dx_0}=\frac{-\xi_\rho}{\xi_0^\pm+\frac{B\xi_\varphi}{\rho^2}}=-\frac{\xi_\rho}{\pm\sqrt{\Delta_1}},
\end{equation}
where  $\Delta_1=\xi_\rho^2+\frac{\xi_\varphi^2}{\rho^2}$.
As in  (\ref{eq:4.5}),  (\ref{eq:4.6}),  (\ref{eq:4.8}),  we have
\begin{equation}																\label{eq:4.26}                
\Delta_2^\pm=\Big(\xi_0^\pm+\frac{B\xi_\varphi}{\rho^2}\Big)^2-\frac{\xi_\varphi^2}{\rho^2},\ \ \xi_\rho^\pm=\mp\sqrt{\Delta_2^+},
\end{equation}  
\begin{equation}																\label{eq:4.27}                 
\frac{d\rho^+}{dx_0}=\frac{\mp\sqrt{\Delta_2^+}}{\sqrt{\Delta_1}},\ \ \ \  \frac{d\rho^-}{dx_0}=\frac{\mp\sqrt{\Delta_2^-}}{-\sqrt{\Delta_1}}.
\end{equation}
The turning points,  i.e.  the solutions of  $\Delta_2^\pm=0$  are  the solutions  of the equations  (\ref{eq:4.15}),  (\ref{eq:4.16})  
with $A=0$.
Therefore   (cf.  (\ref{eq:4.18}))
$$
\frac{1}{\rho^2}=\frac{1}{\rho_0^2}\pm\frac{1}{B|\eta_\varphi|}\pm \frac{1}{B\rho_0}+\dots
$$
 Note that  $|\eta_\varphi|=\rho_0(1-\eta_\rho^2)^{\frac{1}{2}}<\rho_0$.
 
 Hence,  as in  (\ref{eq:4.21}),  (\ref{eq:4.22}),   $\Delta_2^+=0$  for  $\rho_2^+<\rho_1^+<\rho_0$  and  $\Delta_2^-=0$  for 
 $\rho_0<\rho_1^-<\rho_2^-$.
 At  the point  $x_0=0$  we  have $\frac{d\rho^+}{dx_0}=-\eta_\rho>0$   and
 $\frac{d\rho^-}{dx_0}=\frac{-\eta_\rho}{-1}<0$.  Therefore
$\frac{d\rho^+}{dx_0}=\frac{\sqrt{\Delta_2^+}}{\sqrt{\Delta_1}}$  for $x_0>0$,\ 
$\frac{d\rho^-}{dx_0}=  -\frac{\sqrt{\Delta_2^-}}{\sqrt{\Delta_1}}$  for $x_0>0$.    Since  $\Delta_2^+>0$  on  $[\rho_0,+\infty)$ 
we  have that  $\rho^+(x_0)\rw\infty$  when  $x_0\rw +\infty$.  Also   since  $\Delta_2^->0$  for   $\rho<\rho_0$  we have that 
$\rho^-(x_0)\rw 0$  when $x_0$  increases.

It follows  from (\ref{eq:4.26})   that $\Delta_2^-\sim \frac{B^2\xi_\varphi^2}{\rho^4},\ \Delta_1\sim \frac{B^2\xi_\varphi^2}{\rho^4}$.
Therefore
$\frac{d\rho^-}{dx_0}\sim -1$   when  $\rho\rw  0$,  i.e.  $\rho^-(x_0)$  reaches 0 in a finite time  when $x_0$  increases.
Note that 
$$
\frac{d\varphi}{d\rho}=\frac{-\frac{\xi_\varphi}{\rho^2}+(\xi_0^-+\frac{B\xi_\varphi}{\rho^2})\frac{B}{\rho^2}}{\sqrt{\Delta_2^-}}\sim
\frac{\frac{B^2}{\rho^4}\xi_\varphi}{\frac{|B\xi_\varphi|}{\rho^2}}\sim \frac{-B}{\rho^2}.  
$$
Hence  $\varphi\rw \infty$  when  $\rho\rw 0$.  Therefore  $\gamma_-$  approaches  $\rho=0$  spiraling when  $\rho\rw 0$.  

Let  $u^\pm=u_0^\pm+\frac{1}{k}v^\pm$  be  the  same  as  
in Theorem \ref{theo:4.1}  with  $\rho_0,\eta_\varphi,\eta_\rho$  as in (\ref{eq:4.23}),  (\ref{eq:4.24}).
Since $\xi_0^-=-\frac{B\eta_\varphi}{\rho_0^2}-\sqrt{\eta_\rho^2+\frac{\eta_\varphi^2}{\rho_0^2}}>0$
we have  that  $E_0(u^-)<0$  and  $E_{x_0}(u_0^+)>E_0(u^++u^-)$  for large  $k$  as  in  \S 3,  i.e.   the superradiance  take  place.  
Since  
$\rho^+(x_0)\rw \infty$  when  $x_0\rw +\infty$  we have   that  $u_0^+$  escapes  to the  infinity  when $x_0\rw +\infty$.
Also  since  $\rho^-(x_0)\rw  0$  spiraling  when  $x_0$  increases,  we  have that  $u_0^-$  approaches  the singularity  $\rho=0$ 
when  $x_0$  increases.
\qed
\\
\\
{\bf Remark 4.3.  Behavior  of ${\bf u_0^+}$  and ${\bf u_0^-}$  on the time interval (${\bf -\infty,0}$]  in the case of
naked singularity.} 
\\
As in Remark 4.1  we can study first the behavior of $\rho^+(x_0)$  and  $\rho^-(x_0)$  when  $x_0\in (-\infty,0]$.  We  get as in Remark 4.1
that  $\rho^+(x_0)$  decays on $[x_0^{(1)},0]$  where  $x_0^{(1)}<0$.  $x_0^{(1)}$  is the turning point,  i.e.  $\rho^+(x_0^{(1)})=\rho_1^+$.
Then
$\rho^+(x_0)$  increasing  for  $x_0<x_0^{(0)}$  and  tends  to $+\infty$  when  $x_0\rw -\infty$.  Analogously, 
$\rho^-(x_0)$  increases  on  $[x_0^{(2)},0]$  where  $\rho^-(x_0^{(2)})=\rho_1^-$  is the  turning  point  and  then  $\rho^-(x_0)$  decreases  
for  $x_0<x_0^{(2)}$  when  $x_0$  decreases.   Note  that  the  equation  of $\rho^-(x_0)$   for  $x_0<x_0^{(0)}$  is  
$\frac{d\rho^-}{dx_0}=\frac{\sqrt{\Delta_2^-}}{\sqrt{\Delta_1}}$.  Therefore  $\frac{d\rho^-}{dx_0}\sim +1$  when  $\rho\rw 0$
and  $\rho^-(x_0)$  reaches  0 in a finite  time  when  $x_0$  decreases.
Also  $\frac{d\varphi}{d\rho}=O\big(\frac{1}{\rho^2}\big)$.  Thus $\varphi\rw\infty$  when  $\rho\rw 0$,  i.e.  $\rho^-(x_0)$  is approaching  0 spiraling.

Since the null-geodesics  $\gamma_+$  and  $\gamma_-$  are passing  through  the turning points  $\rho_1^+$ and  $\rho_1^-$,
respectively,  we need  the Remark 2.1  to construct  the solutions $u_0^+$  and  $u_0^-$  near turning points.
Since  $\rho^+(x_0)\rw +\infty$  when  $x_0\rw -\infty  $  we get  that   $u_0^+$  escapes  to the  infinity  when  $x_0\rw -\infty. $
Also  $u_0^-$  approaches  the  singularity  $\rho=0$  when  $x_0$  decreases  since  $\rho^-(x_0)\rw 0$  when  $x_0$  decreases.

\section{The acoustic  metric  when  $|B|\leq |A|$}
\label{section 5}
\init

Let $|A|<\rho_0<\sqrt{A^2+B^2},\ A<0, B>0$,  i.e.  $(\rho_0,\varphi)$   belongs  to the  ergoregion.  Suppose
\begin{equation}																\label{eq:5.1}
B<|A|.
\end{equation}
If  $(\eta_\rho,\eta_\varphi)$  are  such  that 
\begin{equation}																\label{eq:5.2}
\lambda^-=-\frac{A\eta_\rho}{\rho_0}-\frac{B\eta_\varphi}{\rho_0^2}-\sqrt{\eta_\rho^2+\frac{\eta_\varphi^2}{\rho_0^2}}>0,
\end{equation}
then $\eta_\rho>0$.   Indeed,
$$
\frac{|B\eta_\varphi|}{\rho_0^2}\leq \frac{B}{|A|}\frac{|\eta_\varphi|}{\rho_0}< \sqrt{\eta_\rho^2+\frac{\eta_\varphi^2}{\rho_0^2}},
$$
and  $\eta_\rho$  must be positive.
\begin{theorem}																	\label{theo:5.1}
Suppose  $0<B\leq |A|$.  
 Choose  $\eta_\rho=\frac{|A|}{\rho_0},\eta_\varphi=-B$.  Let  $u^\pm=u_0^\pm+\frac{1}{k}v^\pm$  be  the same  as in 
Theorem \ref{theo:4.1}  with  the choice  of  $|A|<\rho_0<\sqrt{A^2+B^2},\eta_\rho=\frac{|A|}{\rho_0},\eta_\varphi=-B$.   Then  
$E_0(u_0^-)<0,E_{x_0}(u_0^+)>E_0(u^++u^-)$  for  large  $k$,  i.e.   the superradiance  takes place.  Moreover  
$u_0^+$  and  $u_0^-$  cross the event  horizon  in a finite time.  Therefore  the superradiance is ``short-lived"  outside  the event horizon.
\end{theorem}
{\bf Proof:}
 We have
\begin{equation}																\label{eq:5.3}
\lambda^-=\frac{A^2}{\rho_0^2}+\frac{B^2}{\rho_0^2}-\sqrt{\frac{A^2}{\rho^2}+\frac{B^2}{\rho_0^2}}>0
\end{equation}
since  $\rho_0^2<A^2+B^2$.  It follows from  (\ref{eq:5.3})  that the superradiance  takes place.

As in (\ref{eq:4.8})  we have 
$$
\frac{d\rho^+}{dx_0}=\frac{\pm\sqrt{\Delta_2^+}}{\sqrt {\Delta_1}},\ \ 
\frac{d\rho^-}{dx_0}=\frac{\pm\sqrt{\Delta_2^-}}{-\sqrt {\Delta_1}},
$$
where  $\rho^\pm(x_0)$  are the same as in  (\ref{eq:4.3}),  $\Delta_1$  and  $\Delta_2^\pm$  are the same as  in 
(\ref{eq:4.2}),  (\ref{eq:4.6}).

We shall    show  that  if (\ref{eq:5.1})  holds  then  $\Delta_2^+>0$  for all  $\rho>|A|$,  i.e.  $\rho^+(x_0)$  has no turning points. 
 We get  from  
(\ref{eq:4.16}),  with  $\xi_\varphi=-B$  and  $\xi_0^\pm$  as in  (\ref{eq:5.3})  that 
$$
\delta=4B^4\xi_0^++B^4-4A^2B^2(\xi_0^+)^2=B^2(4B^2\xi_0^++B^2-4A^2(\xi_0^+)^2).
$$  
The roots  of the quadratic equation   $B^2+4B^2t-4A^2t^2=0$  are
$t=\frac{2B^2\pm\sqrt{4B^4+4A^2B^2}}{4A^2}$.  The largest  root $t_1=\frac{B^2+\sqrt{B^4+A^2B^2}}{2A^2}<
\frac{A^2+\sqrt{A^4+A^2A^2}}{2A^2}=\frac{1+\sqrt 2}{2}$.
From other side  $\xi_0^+=\frac{A^2+B^2}{\rho_0^2}+\sqrt{\frac{A^2+B^2}{\rho_0^2}}>1+1=2>\frac{1+\sqrt 2}{2}$.
Thus  $\xi_0^+>t_1$  and consequently  $\delta <0$.  Therefore  $\Delta_2^+>0$  for  all  $\rho$,  i.e.  $\rho^+(x_0)$  has  no turning  points.
Consider   $\frac{d\rho^+}{dx_0}$  when  $x_0=0$.  We have 
\begin{equation}																					\label{eq:5.4}
\frac{d\rho^+}{d x_0}\Big|_{x_0=0}=-\frac{|A|}{\rho_0}-\frac{\frac{|A|}{\rho_0}}{\sqrt{\frac{A^2}{\rho_0^2}+\frac{B^2}{\rho_0^2}}}<0.
\end{equation}
Therefore  $\rho^+(x_0)$  decreases  until  it  reaches  the event  horizon.  
The solution  $\rho^-(x_0)$ can not  leave  the ergoregion  since  $\lambda^-<0$ 
outside  of the  ergoregion  and it 
  always hits  the event  horizon  in finite   time.  
In the all cases  of  the acoustic  metric   $\rho^-(x_0)$  increases  from  $-\infty$
to some point  $x_0^{(2)}$  where 
 it reaches  the turning point $\rho_-$  inside  the ergoregion.  For  $ x_0>x_0^{(2)}$  it decreases  until it reaches  
the event  horizon  (cf.  [4]).  Thus  $u_0^+$  and  $u_0^-$  also  reach  the event  horizon  in  finite time.

Therefore  there is a superradiance  when $B<|A|$  but it  is ``short-lived"  since  both  $u_0^+$  and  $u_0^-$  disappear  inside  the
black  hole after  a finite  time.

\section{The  superradiance  in the case  of white hole}
\label{section 6}
\init

Consider  the acoustic  metric  (\ref{eq:1.8}) when $A>0$. 
 Then  $\rho=A$  will  be a  white hole   (cf [13]).  Note  that the change in the sign  of  $B$  has no  effect  whether  there is  
a black  or white  hole.  From  the other  side  the change  of  the sign of  both  $A$  and  $B$  is equivalent  to the  change  of the sign of
$\xi_0$  in  (\ref{eq:1.8}),  i.e.  to the reversal  of  the time  from  $x_0$  to  $-x_0$.   Therefore  the study  of the  superradiance 
for the case of  the white  hole is equivalent  to the study  of the superradiance  for the black hole on the time interval  $(-\infty,0]$.



\begin{theorem}																			\label{theo:6.1}
Consider  the acoustic  metric  (\ref{eq:1.8})  with  $A>0$.  
Then  $\rho=A$  is a white hole.  Let  $u^\pm=u^\pm+\frac{1}{k}v^\pm$  are  the same  as in  
(\ref{eq:2.10})--(\ref{eq:2.15}).
When the point  $(\rho_0,\varphi_0)$  is  inside  the ergoregion,  $A<\rho_0<\sqrt{A^2+B^2}$
one can choose  $(\eta_\rho,\eta_\varphi)$  such that the superradiance takes  place   (cf. \S 3),   $\mbox{supp}\,u_0^+$  tends  
to the infinity when  $x_0\rw +\infty$   and
$\mbox{supp}\,u_0^-$  tends  
to the event horizon spiraling  when  $x_0\rw +\infty$.
\end{theorem}
{\bf Proof:}  Instead  of giving    
a direct proof  of  Theorem \ref{theo:6.1}  we  change  $x_0$  to  $-x_0$,  i.e.  we consider  the case  of black  hole  on  
$(-\infty,0]$.

 It was shown
in \S 5  that  $\rho^+(x_0)$  with the initial  data $\rho^+(0)=\rho_0$  has  no turning points.  Therefore   since (\ref{eq:5.4})
holds  and  $\Delta_2^+>0$  we  have
$\rho^+(x_0)\rw+\infty$  when $x_0\rw -\infty$.
Also in the case  when  (\ref{eq:4.10})  holds  we have that  $\rho^+(x_0)$  with the initial  data $\rho^+(0)=\rho_0$    has
a  turning  point  $\rho_1^+<\rho_0$.  Therefore  $\rho^+(x_0)$  decreases  when  $x_0$  decreases  until it reaches  
the turning  point $\rho=\rho_1^+$.  After  this  $\rho^+(x_0)\rw+\infty$  when  $x_0\rw -\infty$.  Therefore  in both cases  when
either  (\ref{eq:4.10})  or  (\ref{eq:5.1}) 
 are satisfied  $\rho^+(x_0)\rw +\infty$  when  $x_0\rw -\infty$.  Thus  the support  of corresponding  $u_0^+$  tends to  the infinity  
 when  $x_0\rw -\infty$.
 In both \S 4  and  \S 5  $\rho^-(x_0)$  approaches  the event horizon  spiraling when $x_0\rw -\infty$  and  therefore 
 $\mbox{supp}\, u_0^-$
 also approaches   the event horizon.

Changing  back  $x_0$  to  $-x_0$
we prove  Theorem \ref{theo:6.1}.

\section{The  case of the Kerr metric}
\label{section 7}
\init

In this section we consider  the superradiance  for the case of the Kerr  metric  (\ref{eq:1.2}),  (\ref{eq:1.4}).
Note that $b_z=\frac{z}{r}=0$  when  $z=0$.  Therefore  we have  from  (\ref{eq:2.1})  that  $z\equiv 0$  implies
\begin{equation}																			\label{eq:7.1}
\frac{dz}{ds}=H_{\xi_z}=-2\xi_z+2K(-\xi_0+\hat b\cdot\hat\xi)b_z\equiv 0,
\end{equation}
and therefore  $\xi_z\equiv 0$.

Therefore if the 
initial point  $P_o=(\rho_0,\varphi_0,0)$  is in the equatorial  plane $z=0$  and if $\xi_z=0$  then  the  bicharacteristic system
(\ref{eq:2.1})  is reduced  to the  bicharacteristic system  for  the Hamiltonian
\begin{equation}																			\label{eq:7.2}
H(\rho,\varphi,0,\xi_\rho,\xi_\varphi,0)=\xi_0^2-\xi_\rho^2-\frac{1}{\rho^2}\xi_\varphi^2
+K\Big(-\xi_0+b_\rho\xi_\rho+b_\varphi\frac{\xi_\varphi}{\rho}\Big)^2=0,
\end{equation}
and 
the null-geodesics  lie  in the plane  $z=0$.

As in  (\ref{eq:2.5}),  (\ref{eq:2.6})
\begin{equation}  														\label{eq:7.3}
\lambda^\pm(y_0,\eta')=\frac{Kb'\cdot\eta'\pm\sqrt {\Delta_1}}{1+K}
\end{equation}
 and  
$$
\Delta_1=(1+K)\xi'\cdot\xi'-K(b'\cdot\xi')^2,
$$
where   $z=0,\  b'=(b_\rho,b_\varphi),\  \eta'=(\eta_\rho,\frac{\eta_\varphi}{\rho}),\ y_0=(\rho_0,\varphi_0,0)$.

We will
take  point  $y_0=(\rho_0,\varphi_0,0)$  in the  ergoregion  and choose  $\eta'=(\eta_\rho,\frac{1}{\rho_0}\eta_\varphi)$  such  that
$$
\lambda^-(y_0,\eta')>0.
$$
It follows  from  (\ref{eq:1.4})   for  $z=0$
\begin{equation}																				\label{eq:7.4}
b_\rho=\frac{\sqrt{\rho^2-a^2}}{\rho},\ \ b_\varphi=\frac{a}{\rho},\ \ K=\frac{2m}{\sqrt{\rho^2-a^2}},
\end{equation}
since  $\rho^2=r^2+a^2$   when  $z=0$.  

  We choose
\begin{equation}																\label{eq:7.5}
\eta_\rho=b_\rho(y_0),\ \ \ \eta_\varphi=\rho_0 b_\varphi(y_0),\ \ \ z=0.
\end{equation}
Note that
\begin{equation}                                                								\label{eq:7.6}
\eta_\varphi=\rho_0\frac{a}{\rho_0}=a.
\end{equation}
Since
\begin{equation}																\label{eq:7.7}
b'(y_0)\cdot \eta'=b_\rho^2+b_\varphi^2=1,
\end{equation}																
we have
\begin{align}																\label{eq:7.8}
\Delta_1=&(1+K)(\eta'\cdot\eta')-K( b'\cdot\eta')^2=(1+K)-K=1.
\end{align}
Therefore
\begin{equation}                                  									\label{eq:7.9}
\lambda^\pm(y_0,\eta')=\frac{K\pm 1}{K+1}>0,
\end{equation}
since  $K>1$  in the ergoregion.  Thus  $\lambda^+=1,\ \lambda^-=\frac{K-1}{K+1}<1$.

\begin{theorem}																	\label{eq:7.1}
Consider  the Kerr metric  (\ref{eq:1.1}).  Let  $u^\pm=u_0^\pm+\frac{1}{k}v^\pm$  be the same as in  (\ref{eq:2.10})--(\ref{eq:2.15}).
Choose  initial point  $y_0=(\rho_0,\varphi_0,0)$   in  the ergoregion.   Choose
$\eta_\rho=b_\rho(y_0),\eta_\varphi =\rho_0b_\varphi(y_0),\eta_z=0,z=0$   (cf.  (\ref{eq:7.4})).
Thus  $\lambda^-(y_0,\eta_\rho,\eta_\varphi)=\frac{K-1}{K+1}>0$.
Therefore  the superradiance takes place.  Moreover  $u_0^+$  and  $u_0^-$  disappear  inside  the outer  event horizon  in
a finite  time,  i.e. the superradiance is ``short-lived".
\end{theorem}

We shall study  first  the null-bicharacteristic  (\ref{eq:2.1}) corresponding
to the root  $\xi_0^+=\lambda^+(y_0,\eta')$,  starting  at  $(y_0,\eta')$.  We have
\begin{align}															\label{eq:7.10}
&\frac{d\rho^+}{ds} =\frac{\partial H}{\partial \xi_\rho}=-2\xi_\rho+2K(-\xi_0^++b'\cdot \xi')b_\rho,\ \ \rho^+(0)=\rho_0,
\\																		\label{eq:7.11}
&\frac{d x_0}{ds} =\frac{\partial H}{\partial \xi_0}=2\xi_0^+-2K(-\xi_0^++ b'\cdot \xi'),\ \ x_0(0)=0,
\\																		\label{eq:7.12}
&\frac{d \varphi}{ds} = H_{\xi_\varphi}=  -\frac{2\xi_\varphi}{\rho^2}+2K(-\xi_0^++ b'\cdot \xi') \frac{b_\varphi}{\rho}.
\end{align}

We have  (see  (\ref{eq:2.5}))
\begin{equation}														\label{eq:7.13}
\frac{\partial H}{\partial \xi_0}=2(1+K)\xi_0^+-2K b'\cdot \xi'= 2\sqrt{\Delta_1},
\end{equation}
where  $\Delta_1$  is the same  as in  (\ref{eq:2.6})  for  $z=0,\xi_z=0$.

Also we have
\begin{equation}														\label{eq:7.14}
\frac{\partial  H}{\partial \xi_\rho}  = -2\xi_\rho + 
2K b_\rho(-\xi_0^++ b' \cdot \xi')=2(Kb_\rho^2-1)\xi_\rho+2Kb_\rho\Big(-\xi_0^++b_\varphi\frac{\xi_\varphi}{\rho}\Big).
\end{equation}
  Write  (\ref{eq:7.2})  as the quadratic equation  for $\xi_\rho$:
\begin{equation}														\label{eq:7.15}
(Kb_\rho^2-1)\xi_\rho^2+2K\Big(-\xi_0^++b_\varphi\frac{\xi_\varphi}{\rho}\Big)b_\rho\xi_\rho
+K\Big(-\xi_0^++b_\varphi\frac{\xi_\varphi}{\rho}\Big)^2+(\xi_0^+)^2-\frac{\xi_\varphi^2}{\rho^2}=0.
\end{equation}
Therefore
\begin{equation}														\label{eq:7.16}
\xi_\rho=\frac{-Kb_\rho\Big(-\xi_0^++b_\varphi\frac{ \xi_\varphi}{\rho}\Big)\pm\sqrt{\Delta_2^+}}{Kb_\rho^2-1}
\end{equation}
where
\begin{align}  															\label{eq:7.17}
\Delta_2^+
=&K^2b_\rho^2\Big(-\xi_0^++b_\varphi\frac{\xi_\varphi}{\rho}\Big)^2-
(Kb_\rho^2-1)\Big[K\Big(-\xi_0^++b_\varphi\frac{\xi_\varphi}{\rho}\Big)^2+(\xi_0^+)^2-\frac{\xi_\varphi^2}{\rho^2}\Big]
\\
\nonumber
=&K\Big(-\xi_0^++b_\varphi\frac{\xi_\varphi}{\rho}\Big)^2-(Kb_\rho^2-1)\Big( (\xi_0^+)^2- \frac{\xi_\varphi^2}{\rho^2}\Big).
\end{align}
It  follows  from   (\ref{eq:7.9}),   that  (\ref{eq:7.17})   simplifies to
\begin{equation} 														\label{eq:7.18}
\Delta_2^+=\frac{\rho^2-a^2}{\rho^2}.
\end{equation}

Substituting  (\ref{eq:7.16})  into  (\ref{eq:7.14})  we get
\begin{equation}                 			 								\label{eq:7.19}
\frac{\partial H}{\partial\xi_\rho}=\pm2\sqrt{\Delta_2^+}.
\end{equation}
Therefore 
\begin{equation}														\label{eq:7.20}
\frac{d\rho^+}{dx_0}=\frac{H_{\xi_\rho}}{H_{\xi_0}}=\frac{\pm\sqrt{\Delta_2^+}}{\sqrt{\Delta_1^+}}.
\end{equation}

Thus  $\rho=\rho^+(x_0)$  is either  increasing  or decreasing  
 depending  on the sign  of $\frac{\partial H}{\partial\xi_\rho}$.

It is enough to check the sign  of $\frac{\partial H}{\partial\xi_\rho}$ at the initial  point  $x_0=0$.

Having  $\eta_\rho,\eta_\varphi$  as in  (\ref{eq:7.5})  we get 
\begin{equation}														\label{eq:7.21}
\frac{\partial H}{\partial\xi_\rho}=-2b_\rho(y_0)+2Kb_\rho\Big(-\lambda^+
+b_\rho \eta_\rho
+b_\varphi\frac{\eta_\varphi}{\rho_0}\Big).
\end{equation}
It follows  from  (\ref{eq:7.21})
$$
\frac{\partial H}{\partial \xi_\rho}(y_0,\eta)  = -2b_\rho(y_0)<0,
$$
since  $-\lambda^++b_\rho^2+b_\varphi^2=0$.

Therefore,  $\frac{d\rho^+}{dx_0}<0$  for all  $x_0$  and $\rho=\rho^+(x_0)$  will decay 
when  $x_0$  increases  and will approach  the outer event horizon.

Consider now the ``minus"  null-geodesics  starting  at  $(y_0,\eta')$.   We have as before  
$\frac{d\rho^-}{dx_0}=\pm\frac{\sqrt{\Delta_2^-}}{\sqrt{\Delta_1^-}}$  where $\Delta_2^-$   is   the same as $\Delta_2^+$  
with  $\xi_0^+=1$  replaced  by  $\xi_0^-=\frac{K-1}{K+1}$.

To determine  the sign  of  $\frac{d\rho^-}{dx_0}$  consider  the initial  point  
$y_0=(\rho_0,\varphi_0,0),\ \eta'=(b_\rho,\rho_0 b_\varphi)$.  We have  $\frac{\partial H}{\partial\xi_0}=-2\sqrt{\Delta_1}$
and 
\begin{align}																			\label{eq:7.22}
\frac{\partial H}{\partial\xi_\rho}=-2\eta_\rho+2K b_\rho(-\lambda^-+b'\cdot \eta')
=-2b_\rho+\frac{4K b_\rho}{1+K}>0.
\end{align}
Since  $\frac{\partial H}{\partial \xi_0}<0,\frac{\partial H}{\partial \xi_\rho}>0$   we get that  
\begin{equation}																		\label{eq:7.23}
\frac{d \rho^-}{dx_0}=-\frac{\sqrt{\Delta_2^-}}{\sqrt{\Delta_1}}<0.  
\end{equation}
Therefore  
$\rho^-(x_0)$ also   approaches  the event  horizon  when  $x_0$  increases.

Consider  the behavior  of $\rho^+(x_0)$  and  $\rho^-(x_0)$  when  they  approach  the outer event horizon  $r=r_+$  where  $r=\sqrt{\rho^2-a^2},\ 
r_\pm=m\pm\sqrt{m^2-a^2}$.  Note that  
\begin{align}
\nonumber
&1-Kb_\rho^2=1-\frac{2m}{r}\Big(\frac{r}{\sqrt{r^2+a^2}}\Big)^2
\\
\nonumber
&=\frac{r^2+a^2-2mr}{r^2+a^2}=\frac{(r-r_+)(r-r_-)}{r^2+a^2},
\end{align}
i.e.   $1-Kb_\rho^2=0$  when  $r=r_+$.
We have  $\Delta_2^+\sim K\big(-1+\frac{a^2}{\rho^2}\big)^2$  when  $r\rw r_+$.
Note that (\ref{eq:7.16})  can be  rewritten  in the form  $H_{\xi_\rho}=\pm 2\sqrt{\Delta_2}$.

Since  $H_{\xi_\rho}<0$  for  $\rho^+(x_0)$  at  $x_0=0$  (cf.  (\ref{eq:7.21}))   we take  $-\sqrt{\Delta_2^+}$  in
(\ref{eq:7.16}).  Therefore  when  $\rho\rw \hat\rho_+=\sqrt{a^2+r_+^2}$   we have  canceling of  $\sqrt Kb_\rho-1$:
\begin{multline}																		\label{eq:7.24}
\xi_\rho^+=\frac{-Kb_\rho\big(-1+\frac{a^2}{\rho^2}\big)-\sqrt{\Delta_2^+}}{Kb_\rho^2-1}
\\
\sim\frac{ Kb_\rho\big(1-\frac{a^2}{\rho^2}\big)-\sqrt K\big(1-\frac{a^2}{\rho^2}\big)}
{(\sqrt Kb_\rho-1)(\sqrt K b_\rho+1)}
=\frac{\sqrt K\big(1-\frac{a^2}{\rho^2}\big)}{\sqrt K b_\rho+1}.
\end{multline}
Therefore $\xi_\rho^+$  has a finite  nonzero limit  when  $r\rw r_+$  and therefore  $\frac{d\rho^+}{dx_0}$  has  nonzero limit  
when  $ r\rw r_+$.   Hence  $\rho^+(x_0)$  crosses  the event   horizon  when  $x_0$  increases.  Also  $H_\rho>0$  for  $\rho^-(x_0)$
at  $x_0=0$  (see  (\ref{eq:7.22})).   Therefore   we have   $+\sqrt {\Delta_2^-}$  in  (\ref{eq:7.16})  for  $\rho^-(x_0)$.  Note  that
$\xi_0^-=\frac{K-1}{K+1}=\frac{2m-r_0}{2m+r_0}$ where  $r_0=\sqrt{\rho_0^2-a^2}>r_+$. 
Since  $\frac{2m-r_0}{2m+r_0}<\frac{2m-r_+}{2m+r_+}$  when  $r_+<r_0<2m$  and  $\frac{2m-r_+}{2m+r_+}\leq \frac{a^2}{a^2+r_+^2}$
we get
   that  
$-\xi_0^-+\frac{a^2}{\rho^2}>0$  when  $\rho=\sqrt{a^2+r_+^2}$.   When  $\rho\rw \sqrt{a^2+r_+^2}$
we have 
\begin{equation}																			\label{eq:7.25}
\xi_\rho^-=\frac{-Kb_\rho\big(-\xi_0^-+\frac{a^2}{\rho^2}\big)+\sqrt{\Delta_2^-}}{Kb_\rho^2-1}
\sim 
\frac{-Kb_\rho\big(-\xi_0^-+\frac{a^2}{\rho^2}\big)+\sqrt{K}\big(-\xi_0^-+\frac{a^2}{\rho^2}\big)}{Kb_\rho^2-1}.
\end{equation}
Canceling $-\sqrt Kb_\rho+1$  we  get  that  $\xi_\rho^-$  has a finite nonzero  limit when  $r\rw r_+$.   Therefore  
as in the case of  $\rho^+(x_0)$  we have that $\rho^-(x_0)$  crosses the outer event horizon 
when  $x_0$ increases.  Since  $\rho^+(x_0)$  and $\rho^-(x_0)$ cross the event horizon 
 in a finite   time we get  that  $u_0^+$  and  $u_0^-$  also  cross the outer event  horizon  in a finite  time.   Thus
the superradiance for the Kerr  metric is short-lived  as in \S 5.
\qed
\\
\\
{\bf Remark 7.1. The behavior  of ${\bf u_0^+}$  and ${\bf u_0^-}$  on (${\bf -\infty,0}$].}
\\
Consider the behavior  of $\rho^-(x_0)$  for  $x_0<0$.
Note that  $\rho^-(x_0)$  has a turning  point  $\rho_*,\rho_0<\rho_*<\sqrt{a^2+4m^2},
\Delta_2^-(\rho_*)=0$
because  $\Delta_2^-(\rho_0)>0$  (cf.  (\ref{eq:7.22}))
and  $\Delta_2^-<0$  when   $\hat\rho=\sqrt{4m^2+a^2}$  where  $\rho=\hat\rho$  is the ergosphere.  Indeed,  
$K=1$  on the  ergosphere  and  
\begin{align}																		\label{eq:7.26}
\Delta_2^-
=\Big(-\xi_0^-+\frac{a^2}{\hat \rho^2}\Big)^2+(1-b_\rho^2)\Big((\xi_0^-)^2-\frac{a^2}{\hat \rho^2}\Big)
=(\xi_0^-)^2    -2\xi_0^-\frac{a^2}{\hat \rho^2} +(\xi_0^-)^2\frac{a^2}{\rho^2}.
\end{align}
Thus 
$$
\Delta_2^-=\xi_0^-\Big(1+\frac{a^2}{\hat\rho^2}\Big)\Big[ \xi_0^--\frac{2a^2}{4m^2+a^2}\Big].
$$
We have
$$
\xi_0^-=\frac{2m-r_0}{2m+r_0}<\frac{2m-r_+}{2m+r_+}=\frac{m-\sqrt{m^2-a^2}}{3m +\sqrt{m^2-a^2}}\leq \frac{a^2}{4m^2-a^2}
<\frac{2a^2}{4m^2+a^2}
$$
since  $a<m$.  Hence  $\Delta_2^-<0$  on  the ergosphere.

Note that  $\rho^-(x_0)$  increases  when  $x_0$  decreases  from $0$  to  $x_0^{(1)}$,    
 i.e.  $\rho^-(x_0^{(1)})$   is  the turning 
point.  Then  the sign  of  $\sqrt{\Delta_2^-}$  changes,  i.e.  we change   $+\sqrt{\Delta_+^-}$ to  $-\sqrt{\Delta_+^-}$
in (\ref{eq:7.16}) and the equation  for $\rho^-(x_0)$  becomes  
\begin{equation}																		\label{eq:7.27}
\frac{d\rho^-}{dx_0}=\frac{\sqrt{\Delta_2^-}}{\sqrt{\Delta_1^-}}\ 
\ \ \ \mbox{for}\ \ \ x_0<x_0^{(1)}.
\end{equation}
When  $r\rw r_+$  we  get  $\xi_\rho^-\sim \frac{-2K(-\xi_0^-+\frac{a^2}{\rho^2})}{Kb_\rho^2-1}\sim \frac{1}{r-r_+}$.
Therefore
\begin{equation}																		\label{eq:7.28}
\Delta_1^-=(1+K)\Big((\xi_\rho^-)^2+\frac{a^2}{\rho^2}\Big)-K\Big(b_\rho\xi_\rho^-+\frac{a^2}{\rho^2}\Big)
\sim  \frac{1}{(r-r_+)^2}
\end{equation}
when
$r\rw r_+$,  or  $ \rho\rw \hat\rho_+=\sqrt{r_+^2+a^2}$.   Therefore
\begin{equation}																		\label{eq:7.29}
\frac{d\rho^-}{dx_0}\leq  C|\rho-\hat\rho_+|.
\end{equation}
Hence  $\ln|\rho-\hat\rho_+|\leq Cx_0+C_1$,  and
$$
0<\rho-\hat\rho_+\leq C_2 e^{Cx_0}.
$$
Thus  $\rho\rw \hat \rho_+$  when  $x_0\rw -\infty$.
It follows  from  (\ref{eq:7.12})   that
\begin{equation}																		\label{eq:7.30}
\frac{d\varphi}{d\rho}=O\Big(\frac{1}{r-r_+}\Big).
\end{equation}
Therefore  $\varphi \rw\infty$  when  $\rho\rw\hat\rho_+$.  Thus  the null-geodesics  $\gamma_-$  approaches  the event
horizon  $\rho=\hat\rho_+$  spiraling when  $x_0\rw -\infty$.  
Note that  $\rho^+(x_0)\rw \infty$
when  $x_0\rw -\infty$  since  $\Delta_2^+=\frac{\rho^2-a^2}{\rho^2}>0$  for all  $\rho>a$  and  
$\frac{d\rho^+}{dx_0}=-\frac{\sqrt{\Delta_2^+}}{\sqrt{\Delta_1^+}}<0$.
\\
\\
{\bf  Remark  7.2.  }
\\
  The case  of  the  white  Kerr hole arises  when  the time  direction  is reversed  (cf.  \S 6).
It is  equivalent  to studying  the
black Kerr hole  on  $(-\infty,0]$.   This study was done in Remark 7.1.
 
Thus we have that $\mbox{supp}\,u_0^+$  tends  to  infinity  when  $x_0\rw -\infty$.  Since  $\gamma_-$  spirals  approaching
$\rho=\hat\rho_+$  when  $x_0\rw -\infty$,  we have  that  $\mbox{supp}\,u_0^-$  
also  approaches  $\rho=\hat \rho_+$  when  $x_0\rw -\infty$.
\begin{figure}
\centering

\begin{tikzpicture}[scale=0.4]

\draw(0,0) circle  [radius = 2]; 
\draw(0,0) circle  [radius = 5.5];
\draw(0,0) circle  [radius = 8];

\draw(0,0) circle  [radius = 11];

\draw[<-][line width = 3 pt](1.4,1.35) .. controls   (3,4)      and (3.5,6) ..(3,9);
\draw[line width = 3 pt](3,9) .. controls (1.5,12)  ..(-1,15);
\draw[<-][line width = 3 pt] (-1,15) -- (-2,16);

\draw[<-][line width = 1 pt,double distance= 1pt](2,0) .. controls (5,3) and (6,8).. (3,9);
\draw[line width = 1 pt,double distance = 1pt](3,9) .. controls (-1,12) and (-8,8) ..  (-8.7,0);
\filldraw[black] (-0.9,10 ) circle (8pt);
\filldraw[black] (3,9) circle (8pt);

\draw(-1,-2.6) node {$\rho=a$};
\draw(1,-6.2) node {$\rho=\sqrt{a^2+r_-^2}$};
\draw(3,-8.3) node {$\rho=\sqrt{a^2+r_+^2}$};
\draw(4,-11.3) node {$\rho=\sqrt{a^2+4m^2}$};

\draw(3.8,9.5)  node {$P_0$};
\draw(2.4,12) node {$\gamma_+$};
\draw(-2.4,8.7)  node  {$\gamma_-$};

\end{tikzpicture}

\caption{\label{myfigure-2}  The case $a<m$.}  
 Null-geodesics  $\gamma_+$  starts  at  $P_0$  
and  reaches  $\rho=a$  at a finite  time  $x_0^{(1)}>0$.
When  $x_0\rw -\infty$  $\gamma_+$  tends  to   infinity.
Null-geodesics   $\gamma_-$  starts  at $P_0$   and reaches  $\rho=a$  at  finite time $x_0^{(2)}>0$.
When  $x_0<0$  $\gamma_-$ passes a turning point and then  spirals, approaching  the outer event  horizon.
\end{figure}
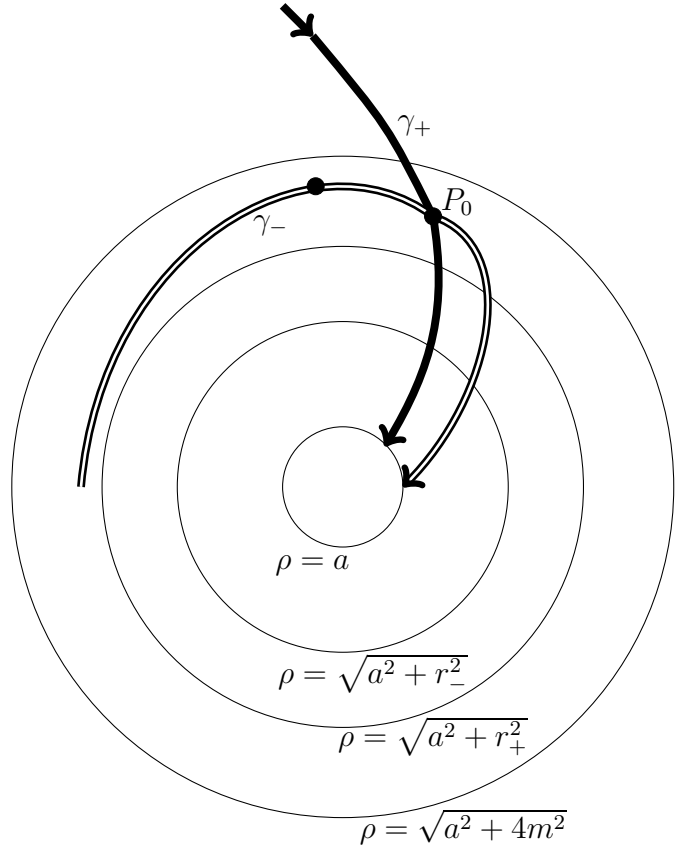

\section{The behavior of solutions inside   the Kerr outer horizon}
\label{section 8}
\init

\subsection{The case of null-geodesics  in the equatorial  plane}

Consider what  happens  with  the  ``plus"  and  ``minus"  null-geodesics  $\gamma_+$  and  $\gamma_-$  after they  cross 
the  outer event horizon  $r_+=m+\sqrt{m^2-a^2}$.  

\begin{theorem}																		\label{theo:8.1}
After crossing  the outer  event  horizon   both  null-geodesics $\gamma^+$  and  $\gamma^-$
cross also  the inner  event  horizon  and end  on  the  singularity  ring  $\rho=a$    at a finite time  (see Fig. 1).
\end{theorem}
 Consider  first the case   of the  ``plus"   null-geodesics. We have,  as in  \S 7:
\begin{equation}																	\label{eq:8.1}
\frac{d\rho^+(x_0)}{dx_0}=-\frac{\sqrt{\Delta_2^+}}{\sqrt{\Delta_1^+}},
\end{equation}
where $\Delta_2^+=\frac{\rho^2-a^2}{\rho^2},$
\begin{equation}																	\label{eq:8.2}
\Delta_1^+=(1+K)\Big((\xi_\rho^+)^2+\frac{\xi_\varphi^2}{\rho^2}\Big)-K\Big(b_\rho\xi_\rho^++b_\varphi\frac{\xi_\varphi}{\rho}\Big)^2.
\end{equation}
Since
$\Delta_2^+\neq 0, \Delta_1^+\neq  0$  we get  from  (\ref{eq:8.1})   that $\rho^+(x_0)$  decreases 
when  $x_0$  increases.  After  $\gamma_+$  cross the outer  event  horizon  it crosses  also  the inner  event  horizon  
since (\ref{eq:7.24})  holds  near  $r=r_0$.  Then it
is 
approaching the ring  singularity  $(\rho-a)^2+z^2=0$

We  shall study  the behavior of  $\gamma_+$  near  the ring singularity,  i.e.  when $z=0,\ \rho$  is close  to $a,\ \rho>a$.
It follows from (\ref{eq:7.16}) that
\begin{equation}																	\label{eq:8.3}
\xi_\rho^+=\frac {-\frac{2m}{\sqrt{\rho^2-a^2}}\frac{\sqrt{\rho^2-a^2}}{\rho}\big(-1+\frac{a^2}{\rho^2}\big)+ \frac{\sqrt{\rho^2-a^2}}{\rho}}
{\frac{2m}{\sqrt{\rho^2-a^2}}\big(\frac{\sqrt{\rho^2-a^2}}{\rho}\big)^2-1}.
\end{equation}
Hence 
\begin{multline}																	\label{eq:8.4}
\Delta_1^+=\Big(1+\frac{2m}{\sqrt{\rho^2-a^2}}\Big)(1+O(\rho^2-a^2))-\frac{2m}{\sqrt{\rho^2-a^2}}(1+O(\rho^2-a^2))
\\
=
1+O(\sqrt{\rho^2-a^2}).
\end{multline}
Thus
$
\frac{d\rho^+}{dx_0}=-\frac{\sqrt{\Delta_2^+}}{\sqrt{\Delta_1^+}}\leq -C\sqrt{\rho-a}.
$
Therefore
\begin{equation}																		\label{eq:8.5}
\frac{d\rho}{\sqrt{\rho-a}}\leq -C dx_0.
\end{equation}
Integrating (\ref{eq:8.5})  from  $x_0$  to  $t$,  where $x_0<t,\ \rho^+(t)-a>0$,  we get
\begin{equation}																		\label{eq:8.6}
2\sqrt{\rho^+(t)-t}-2\sqrt{\rho^+(x_0)-a}\leq -C(t-x_0).
\end{equation}
Increasing $t$  further  we get  $t_0$  such  that $\rho^+(t_0)-a=0$,  i.e.  $\rho^+(x_0)$  reaches  $\rho=a$  at  $x_0=t_0$.

Note that  (cf. (\ref{eq:7.12})
\begin{equation}																		\label{eq:8.7}
\frac{d\varphi}{d\rho}=\frac{H_{\xi_\varphi}}{H_{\xi_\rho}}=
\frac{-\frac{2\xi_\varphi}{\rho^2}+2K\big(-\xi_0^++b_\rho\xi_\rho^++b_\varphi\frac{\xi_\varphi}{\rho}\big)\frac{b_\varphi}{\rho}}
{\pm\sqrt{\Delta_2^+}}
=\frac{C(\rho)}{(\rho^2-a^2)^{\frac{1}{2}}},
\end{equation} 
where  $C(\rho)\neq 0$. 
  Hence $\varphi(x_0)\rw\varphi_0$   when  $\rho\rw a$.  Therefore  the  ``plus"  null-geodesics  $\gamma_+$   ends  at the point  
  $(a,\varphi_0)$  of the singularity ring  at some time  $t_0$.
  
  Now consider the case  of  ``minus"  null-geodesics  $\gamma_-$.  It also crosses  the inner  event horizon when  $x_0$ 
  increases since  (\ref{eq:7.25})   holds.   Note that  $0<\xi_0^-=\lambda^-<1$   (cf.  (\ref{eq:7.9})).  We have  (cf.  (\ref{eq:7.17}))
  \begin{align}																			\label{eq:8.8}
  \Delta_2^-&= K\Big(-\xi_0^-+\frac{\rho^2}{a^2}\Big)^2-(Kb_\rho^2-1)\Big((\xi_0^-)^2-\frac{a^2}{\rho^2}\Big)
  \\
  \nonumber
  &=\frac{2m}{\sqrt{\rho^2-a^2}}\Big((-\xi_0^-+1)+\frac{\rho^2-a^2}{a^2}\Big)^2
  \\
  \nonumber
  &-\Big(\frac{2m}{\sqrt{\rho^2-a^2}}\Big(\frac{\sqrt{\rho^2-a^2}}{\rho}\Big)^2-1\Big)\Big((\xi_0^-)^2-1+\frac{\rho^2-a^2}{a^2}\Big)^2
  \\
  \nonumber
  &=
  \frac{2m(1-\xi_0^-)^2}{\sqrt{\rho^2-a^2}}(1+O(\sqrt{\rho^2-a^2})).
  \end{align}
  Also we have
  (cf.  (\ref{eq:7.16})
  \begin{multline}																\label{eq:8.9}
  \xi_\rho^-=\frac{-Kb_\rho\big(-\xi_0^-+\frac{a^2}{\rho^2}\big)+\sqrt{\Delta_2^+}}{Kb_\rho^2-1}=
  +\sqrt{\Delta_2^-}
  +O(1).
  \end{multline}
Hence  (cf.  (\ref{eq:8.2})
\begin{multline}																	\label{eq:8.10}
\Delta_1^-=\Big(1+\frac{2m}{\sqrt{\rho^2-a^2}}\Big)\Big(\Big(\xi_\rho^-\Big)^2+\frac{a^2}{\rho^2}\Big)
-\frac{2m}{\sqrt{\rho^2-a^2}}\Big(\frac{\sqrt{\rho^2-a^2}}{\rho}\xi_\rho^-+\frac{a^2}{\rho^2}\Big)^2
\\
=\frac{(2m)^2(1-\xi_0^-)^2}{\rho^2-a^2}+O\Big(\frac{1}{\sqrt{\rho^2-a^2}}\Big).
\end{multline}
It follows  from  (\ref{eq:7.22})   and  (\ref{eq:8.8}),  (\ref{eq:8.10})  that
\begin{equation} 																	\label{eq:8.11}
\frac{d\rho^-}{dx_0}=-\frac{\sqrt{\Delta_2^-}}{\sqrt{\Delta_1^-}}
\leq -\frac{C_1(\rho^2-a^2)^{-\frac{1}{4}}}{C_2(\rho^2-a^2)^{-\frac{1}{2}}}
\leq-C_4(\rho-a)^{\frac{1}{4}},
\end{equation}
i.e.
$$
(\rho-a)^{-\frac{1}{4}} d\rho
\leq -C_4 dx_0.
$$
Analogously to (\ref{eq:8.6}),   integrating from  $x_0$  to $t_1, \ t_1> x_0, \ \rho^-(t_1)-a>0,$   
we  get that there exists  $t^{(0)}$  such that  $\rho^-(t^{(0)})-a=0$,  i.e.   $\gamma_-$  also  ends  on  $\rho=a$.

\subsection{Behavior  of  ${\bf u_0^+}$ and  ${\bf u_0^-}$  inside  the outer horizon}

Note  that  $P_0=(\rho_0,\varphi_0,0)$  and the null-geodesics  $\gamma_-,\gamma_+$  starting  at $P_0$  are lying in
the equatorial plane $z=0$,  but  $U_0$  is a neighborhood  of  $P_0$  in $(\rho,\varphi,z)$  space.  Therefore  the null-geodesics  
starting on $\mbox{supp}\,\chi_0$  are lying not  only
 in  the  equatorial  plane  and the solutions $u^\pm$  constructed  in \S 2 are  the  solutions  of  the  equation (\ref{eq:2.7})  in
 the three-dimensional  space.

\begin{theorem}																		\label{theo:8.2}
Both  
 $u_0^+$ and  $u_0^-$  cross the outer  and the inner   horizons  and 
both end on the singularity ring $(\rho-a)^2+z^2=0$. 
\end{theorem}
{\bf Proof:} To prove   this we shall  show  that all ``plus"  and ``minus"  
null-geodesics starting  on $\mbox{supp}\,\chi_0$ will end  on  $(\rho-a)^2+z^2=0$.   
Thus we need to estimate the behavior of null-geodesics  in $\R^3$
starting  on $\mbox{supp}\,\chi_0$ and this is more difficult than the estimates for the null-geodesics  lying in the equatorial  plane.

We shall start  with  $u_0^+$  and  will describe the behavior of plus null-geodesics
$\gamma_+^{(1)}$ starting  on $\mbox{supp}\,\chi_0$  and close
to $\gamma^+$,  i.e.  $\gamma_+^{(1)}$  is  the projecion   on  $(\rho,\varphi,z)$ of the null-bicharacteristic,  with
the initial data $(\rho',\varphi',z',\eta_\rho,\eta_\varphi,\eta_z)$ where  $(\rho',\varphi',z')$  are  cloze  to  $(\rho_0,\varphi_0,0)$ 
 and  $(\eta_\rho,\eta_\varphi,\eta_z)$  are the same  as for  $\gamma^+$.  Note that $z'$  is small but not equal  to zero.  
 Suppose,  for the definiteness,  that  $z'>0$.
 
 We have  the same formulas as in \S 7  and  \S 8.1,  with the variables $z,\xi_z$ added:
\begin{equation}														\label{eq:8.12}
\frac{d\rho^+}{dx_0}=-\frac{\sqrt{\Delta_2^+}}{\sqrt{\Delta_1^+}},\ \ \
\frac{d z}{dx_0}=-\frac{\sqrt{\Delta_3^+}}{\sqrt{\Delta_1^+}},
\end{equation}
where 
\begin{equation}														\label{eq:8.13}
\Delta_1^+=(1+K)\Big(\xi_\rho^2+\xi_z^2+\frac{\xi_\varphi^2}{\rho^2}\Big)
-K\Big(b_\rho\xi_\rho+b_z\xi_z+b_\varphi\frac{\xi_\varphi}{\rho}\Big)^2.
\end{equation}
Now  $K=\frac{2mr^3}{r^4+a^2z^2}$   where  $r$   is defined as in (\ref{eq:1.5}). Also
\begin{equation}														\label{eq:8.14}
\Delta_2^+=K (-\xi_0^++b_z\xi_z+b_\varphi\frac{\xi_\varphi}{\rho})^2-
(Kb_\rho^2-1)\Big((\xi_0^+)^2-\xi_z^2-\frac{\xi_\varphi^2}{\rho^2}\Big),
\end{equation}
\begin{equation}														\label{eq:8.15}
\Delta_3^+=K (-\xi_0^++b_\rho\xi_\rho+b_\varphi\frac{\xi_\varphi}{\rho})^2+
(Kb_z^2-1)\Big((\xi_0^+)^2-\xi_\rho^2-\frac{\xi_\varphi^2}{\rho^2}\Big).
\end{equation}
Near the singularity ring   $(\rho -a)^2 +z^2=0$ we have  from  (\ref{eq:1.5})
\begin{equation} 														\label{eq:8.16}
C_1'((|\rho-a|+|z|)\leq C_1\sqrt{(\rho-a)^2+z^2}
 \leq r^2
  \leq C_2 \sqrt{(\rho-a)^2+z^2}
  \leq C_2(|\rho-a|+|z|),
\end{equation}
i.e.   $r\sim (|\rho-a|+|z|)^{\frac{1}{2}}$  where 
$\alpha\sim\beta$  means  $C_1\beta\leq \alpha\leq C_2\beta$.

Therefore  from  (\ref{eq:1.4})  and  (\ref{eq:8.16})  we get
\begin{equation}														\label{eq:8.17}
K\sim\frac{(|\rho-a|+|z|)^{\frac{3}{2}}}{(|\rho-a|+|z|)^2+a^2z^2}\sim \frac{1}{(|\rho-a|+|z|)^{\frac{1}{2}}},
\end{equation}
\begin{align}														\label{eq:8.18}
&b_\rho = \frac{r\rho}{r^2+a^2}\sim (|\rho-a|+|z|)^{\frac{1}{2}},\ \ \ b_z=\frac{z}{r}\sim\frac{z}{(|\rho-a|+|z|)^{\frac{1}{2}}},
\\
\nonumber
&
\frac{\xi_\varphi}{\rho}=\frac{a}{\rho}\sim 1,\ \  b_\varphi = \frac{a\rho}{r^2+a^2}\sim 1,\ \ 
b_\varphi\frac{\xi_\varphi}{\rho}=\frac{a^2}{r^2+a^2}\sim 1-O(r^2).
\end{align}
We  shall estimate  $\xi_\rho^2$  and $\xi_z^2$  from above.  It follos  from  (\ref{eq:1.2})  with $H=0$   that
\begin{equation}														\label{eq:8.19}
\xi_\rho^2+\xi_z^2=K\Big(b_\rho\xi_\rho+b_z\xi_z-\xi_0^+  +b_\varphi\frac{\xi_\varphi}{\rho}\Big)^2
+(\xi_0^+)^2-\frac{\xi_\varphi^2}{\rho^2}.
\end{equation}
Therefore
\begin{multline}					
\nonumber
\xi_\rho^2+\xi_z^2\leq K\Big(-\xi_0^++b_\varphi\frac{a}{\rho}+b_z\xi_z\Big)^2
+2K \Big|-\xi_0^++b_\varphi\frac{a}{\rho}+b_z\xi_z\Big|b_\rho|\xi_\rho|
\\
+Kb_\rho^2|\xi_\rho|^2 
+\frac{\rho^2-a^2}{\rho^2}.
\end{multline}
Using   the identity  $2cd =\e^2c^2+\frac{d^2}{\e^2}-\big(\e c -\frac{d}{\e}\big)^2$  and  choosing  $\e^2=\frac{Kb_\rho^2}{1-Kb_\rho^2},
\ d=\sqrt K b_\rho|\xi_\rho|, \ c=\sqrt K\big|-\xi_0^++b_\varphi\frac{a}{\rho}+b_z\xi_z\big|,$
we get
\begin{align}																		\label{eq:8.20}
&\xi_\rho^2+\xi_z^2\leq \frac{1}{1-Kb_\rho^2}K\Big(-\xi_0^++b_\varphi\frac{a}{\rho}+b_z\xi_z\Big)^2
+\xi_\rho^2+\frac{\rho^2-a^2}{\rho^2}
\\ 
\nonumber
-&\Big|\e\sqrt K \big|-\xi_0^++b_\varphi\frac{a}{\rho}+b_z\xi_z\big|
-\frac{\sqrt K}{\e}b_\rho|\xi_\rho| \Big|^2.
\end{align}
Canceling $|\xi_\rho^2|$  in  (\ref{eq:8.20})  we get
\begin{equation}																	\label{eq:8.21}
|\xi_z|^2\leq  \frac{1}{1-Kb_\varphi^2}K\Big(-\xi_0^++b_\varphi\frac{a}{\rho}+b_z\xi_z\Big)^2+\frac{\rho^2-a^2}{\rho^2}-I_1^2,
\end{equation}
where
 \begin{equation}																	\label{eq:8.22}
 I_1=|\e\sqrt K|-\xi_0^++b_\varphi\frac{a}{\rho}+b_z\xi_z |-\frac{\sqrt K}{\e}b_\rho | \xi_\rho|\, \big|,\ \ 
\e^2=\frac{Kb_\rho^2}{1-Kb_\rho^2}.
\end{equation}
Using  (\ref{eq:8.21})  we have
\begin{align}																		\label{eq:8.23}
&\Delta_2^+=K\Big(-\xi_0^++b_\varphi\frac{a}{\rho}+b_z\xi_z\Big)^2+(1-Kb_\rho^2)\Big(\frac{\rho^2-a^2}{\rho^2}-\xi_z^2\Big)
\\
\nonumber
&\geq 
K\Big(-\xi_0^++b_\varphi\frac{a}{\rho}+b_z\xi_z\Big)^2
\\
\nonumber
&+(1-Kb_\rho^2)\Big(\frac{\rho^2-a^2}{\rho^2}-
\frac{1}{1-Kb_\rho^2}  K\Big(-\xi_0^++b_\varphi\frac{a}{\rho}+b_z\xi_z\Big)^2  -\frac{\rho^2-a^2}{\rho^2}+I_1^2\Big)
\\
\nonumber
&=(1-Kb_\rho^2) I_1^2 \geq I_1^2. 
\end{align}		
Analogously  we have,  changing  $b_\rho$  to  $b_z$  and  $|\xi_\rho|$  to  $|\xi_z|$
\begin{equation}																\label{eq:8.24}
\Delta_3^+\geq (1-Kb_z^2)I_2^2\geq I_2^2,
\end{equation}
where  
\begin{equation}																	\label{eq:8.25}
I_2=\big|\e_1\sqrt K\big |-\xi_0^++b_\varphi\frac{a}{\rho}+b_\rho\xi_\rho|-\frac{\sqrt K}{\e_1}|b_z|\,|\xi_z|\big|,
\end{equation}
where $\e_1=\frac{Kb_z^2}{1-Kb_z^2}$.

Now estimate  $\Delta_1^+$.  We have  
$$
\Delta_1^+=(1+K)\Big(\xi_\rho^2+\xi_z^2+\frac{a^2}{\rho^2}\Big)
-K\Big(b_\rho\xi_\rho+b_z\xi_z+b_\varphi\frac{a}{\rho}\Big)^2.
$$
It follows  from  (\ref{eq:8.19})  that
\begin{align}                    		 													\label{eq:8.26}
&\xi_\rho^2+\xi_z^2\leq 2K\Big(-\xi_0^++b_\varphi\frac{a}{\rho}\Big)^2  +2K(b_\rho\xi_\rho+b_z\xi_z)^2 +\frac{\rho^2-a^2}{\rho^2}
\\
\nonumber
&\leq 2K\Big(-1+\frac{a^2}{r^2+a^2}\Big)^2 +2K(b_\rho^2+b_z^2)(\xi_\rho^2+\xi_z^2)+\frac{\rho^2-a^2}{\rho^2}.
\end{align}
Since  $b_\rho^2+b_z^2=1-b_\varphi^2=1-\frac{a^2\rho^2}{(r^2+a^2)^2}=O(r^2)$  we get   from  (\ref{eq:8.17}),  (\ref{eq:8.18})
\begin{equation}																	\label{eq:8.27}
\xi_\rho^2+\xi_z^2\leq (1-2K(b_\rho^2+b_z^2))^{-1}\Big(\frac{\rho^2-a^2}{\rho^2}+C\delta^{\frac{3}{2}}\Big)
\leq C_1\Big(\frac{\rho^2-a^2}{\rho^2}+C\delta^{\frac{3}{2}}\Big),
\end{equation}
where
\begin{equation}																	\label{eq:8.28}
\delta=|\rho-a|+|z|.
\end{equation}
Since  $K\frac{a^2}{\rho^2}-Kb_\varphi^2\frac{a^2}{\rho^2}=O(\delta^{\frac{1}{2}})$,  we get,  using  (\ref{eq:8.27})
\begin{equation}																	\label{eq:8.29}
1-C\delta^{\frac{1}{2}}<\Delta_1^+ < 1+C\delta^{\frac{1}{2}}.
\end{equation}  
We will need  also  the estimate  of  $\xi_\rho^2+\xi_z^2$  from below.  It follows  from  (\ref{eq:8.19}),
using  the  inequality of  the form $(c+d)^2\geq \frac{c^2}{2}-d^2$,   that
\begin{align}				
\nonumber
&\xi_\rho^2+\xi_z^2\geq \frac{K}{2}\Big(-\xi_0^++b_\varphi\frac{a}{\rho}\Big)^2-K(b_\rho\xi_\rho+b_z\xi_z)^2+\frac{\rho^2-a^2}{\rho^2}
\\
\nonumber
&\geq\frac{K}{2}\Big(-\xi_0^++b_\varphi\frac{a}{\rho}\Big)^2-K(b_\rho^2+b_z^2)(\xi_\rho^2+\xi_2^2)
+\frac{\rho^2-a^2}{\rho^2}.
\end{align}
Therefore 
\begin{equation}																	\label{eq:8.30}
\xi_\rho^2+\xi_2^2\geq \Big(1+ K(b_\rho^2+b_z^2)\Big)^{-1}\Big(
\frac{K}{2}\big(-\xi_0^++b_\varphi\frac{a}{\rho}\big)^2   
+\frac{\rho^2-a^2}{\rho^2}\Big).
\end{equation}
Using  (\ref{eq:8.17}),  (\ref{eq:8.18})  we got  
\begin{equation} 																	\label{eq:8.31}
\xi_\rho^2+\xi_z^2\geq C\big(|\rho-a|+\delta^{\frac{3}{2}}\big)\geq  C\delta^{\frac{3}{2}}.
\end{equation}

Let  $(\rho',\varphi',z',\eta_\rho,\eta_\varphi,\eta_z)$  be the  initial  data for  the null-bicharacteristic system  (\ref{eq:2.1}).
Assuming  that there are  no  focal  points,   there is a one-to-one correspondence  
between $(\rho(x_0),z(x_0))$  and $(\rho',z')$.
Therefore 
$z(x_0)>0$  if  $z(0)=z'>0$.
Hence      
we shall  write  $\delta(x_0)=\rho(x_0)-a+z(x_0)$  instead  of   $\delta=|\rho(x_0)-\alpha|+|z|$  since  $z(x_0)>0$  and  $\rho(x_0)-a>0$.

We have  from  (\ref{eq:8.12}), (\ref{eq:8.23}), (\ref{eq:8.24}), (\ref{eq:8.29})  that
\begin{equation}																	\label{eq:8.32}
\frac{d\delta}{dx_0}=\frac{d\rho}{dx_0}+\frac{dz}{dx_0}=-\frac{\sqrt{\Delta_2^+}}{\sqrt{\Delta_1^+}}-
\frac{\sqrt{\Delta_3^+}}{\sqrt{\Delta_1^+}}\leq -C(I_1+I_2),
\end{equation}
where  $\delta =\rho-a+z$.

Note that  (cf.  (\ref{eq:8.17}),  (\ref{eq:8.18}))
\begin{align}																							\label{eq:8.33}
&
I_1=\Big|\frac{K^{\frac{1}{2}}b_\rho K^{\frac{1}{2}}} {(1-Kb_\rho^2)^{\frac{1}{2}}} \big|-\xi_0^++b_\varphi\frac{a}{\rho}+b_z\xi_z\big|
-\frac{\sqrt K (1-Kb_\rho^2)^{\frac{1}{2}}b_\rho|\xi_\rho|} 
{\sqrt K|b_\rho|}
\Big|
\\
\nonumber
\geq
&|1-Kb_\rho^2|^{\frac{1}{2}}|\xi_\rho|-C_1\delta   \geq  C|\xi_\rho|-C_1\delta.
\end{align}
Analogously 
\begin{equation}																					\label{eq:8.34}
I_2\geq C|\xi_z|-C_2\delta.
\end{equation}
Using  the estimate  (\ref{eq:8.31})   we get
\begin{equation}                 																		\label{eq:8.35}
\frac{d\delta}{d x_0}\leq -C(I_1+I_2)\leq -C(|\xi_\rho|+|\xi_z|+C_3\delta)\leq  -C\delta^{\frac{3}{4}}+C_1\delta\leq -C_2\delta^{\frac{3}{4}}.
\end{equation}
As  in  (\ref{eq:8.6})   we conclude  from  (\ref{eq:8.35})  that    $\delta(t_0)=0$  for  some  $t_0$.   Therefore 
all null-geodesics  $\gamma_+^{(1)}$  end   on the singularity  ring $ (\rho-a)^2+z^2=0$.  
   
Now we shall study the behavior of  null-geodesics $\gamma_-^{(1)}$ close  to  $\gamma^-$
 inside the outer  event  horizon.  Note that the case of 
$\gamma^-$  null-geodesics differ  from the case  of $\gamma^+$  null-geodesics  only  by $\xi_0^-=\frac{K_0-1}{K_0+1}$
replacing  $\xi_0^+=1$.  Here $K_0$  is  the $K=\frac{2mr^3}{r^4+a^2z^2}$  at the initial  point $(\rho_0,\varphi_0,0)$.
In particular,  
instead  of  (\ref{eq:8.23}),  (\ref{eq:8.24})  we  have
\begin{equation}             																			\label{eq:8.36}
\Delta_2^-\geq  I_3^2,\ \ \ \Delta_3^-\geq I_4^2,
\end{equation}
where
\begin{align}																					\label{eq:8.37}
&I_3=\Big|\e\sqrt K |-\xi_0^-+b_\varphi\frac{a}{\rho}+b_z\xi_z|-\frac{\sqrt K}{\e}|b_\rho|\, |\xi_\rho|\Big|,\ \ 
\e^2=\frac{Kb_\rho^2}{1-Kb_\rho^2},
\\
\nonumber
&I_4=\Big|\e_1\sqrt K |-\xi_0^-+b_\varphi\frac{a}{\rho}+b_\rho\xi_\rho|-\frac{\sqrt K}{\e_1}|b_z|\, |\xi_z|\Big|,\ \ 
\e_1^2=\frac{Kb_z^2}{1-Kb_z^2}.
\end{align}
Estimate  (\ref{eq:8.27})  takes  the form
\begin{equation}																				\label{eq:8.38}
\xi_\rho^2+\xi_z^2\leq CK\Big(-\xi_0^-+\frac{a^2}{r^2+a^2}\Big)^2  +C\big|-\xi_0^-+\frac{a^2}{\rho^2}\big|\leq C\delta^{-\frac{1}{2}}..
\end{equation}
The estimate  of  $\xi_\rho^2+\xi_z^2$  from  below  have  the form  (cf.  (\ref{eq:8.30}))
\begin{equation}																				\label{eq:8.39}
\xi_\rho^2+\xi_z^2\geq  \frac{C}{\delta^{\frac{1}{2}}}.
\end{equation}
Combining  (\ref{eq:8.38})  and  (\ref{eq:8.39})  we  get  instead of  (\ref{eq:8.29})
\begin{equation}																				\label{eq:8.40}
C_1\delta^{-1}\leq  \Delta_1^-\leq  C_2\delta^{-1}.
\end{equation}
Therefore it follows from  (\ref{eq:8.32}),  (\ref{eq:8.40}) 
\begin{equation}																				\label{eq:8.41}
\frac{d\delta}{dx_0}=\frac{d\rho}{dx_0}+\frac{dz}{dx_0}\leq -\frac{C(I_3+I_4)}{\delta^{-\frac{1}{2}}}.
\end{equation}
We have  similarly  to (\ref{eq:8.33})
\begin{equation}																				\label{eq:8.42}
I_3\geq  C|\xi_\rho|-C_1,\ \ I_4\geq  C|\xi_z|-C_1.
\end{equation}
Therefore  
\begin{equation}																				\label{eq:8.43}
\frac{d\delta}{dx_0}\leq  -C\delta^{\frac{1}{2}}(|\xi_\rho|+|\xi_z|-C_1)
\leq  -C\delta^{\frac{1}{2}}\delta^{-\frac{1}{4}}=-C\delta^{\frac{1}{4}}.
\end{equation}
From  (\ref{eq:8.43})  analogously to  (\ref{eq:8.11})  we get  that  $\delta=\rho^--a+z^-$ vanish  at the singularity  ring.
Therefore  all null-geodesics  $\gamma_-^{(1)}$ end on the  singularity ring  (see Fig. 1).

Hence  both  $u_0^+$  and  $u_0^-$  end on  the singularity ring  at a finite  time.

\section{The superradiance  in the case  of  extremal ${\bf (a^2=m^2)}$  Kerr metric and 
  naked singularity  ${\bf (a^2>m^2)}$  Kerr matric}
\label{section 9}
\init

In both cases the construction of the solutions $u^+$  and  $u^-$  remains 
the  same and  we  only  indicate  the difference  in  the behavior  of  the null-geodesics  $\gamma_+$  and  $\gamma_-$.

In the case  of extremal metric  $a^2=m^2$  we have  $r_+=r_-=m$,  i.e  the outer  and  the inner horizons coincide.
As in the case $a<m$   $\rho^+(x_0)$  and  $\rho^-(x_0)$  cross the event  horizon  when $x_0$  increases.

Consider  the behavior  of $\rho^-(x_0)$  when $x_0$  decreases.  Note that  $Kb_\rho^2-1=O((r-r_+)^2)$  and therefore  
(cf.  (\ref{eq:7.27})):
\begin{equation}           															\label{eq:9.1}
\xi_\rho^-\sim \frac{1}{(r-r_+)^2}.
\end{equation}
Therefore $\sqrt{\Delta_1^-}\sim \frac{1}{(r-r_+)^2}$  and,  similarly to  (\ref{eq:7.29}),
\begin{equation}																	\label{eq:9.2}
\frac{d\rho^-}{dx_0}\leq C_0(\rho-\hat\rho_+)^2,  \ \  C_0>0,
\end{equation}
where $\hat \rho_+=\sqrt{a^2+m^2}$.

Hence  $-\frac{1}{\rho^--\hat \rho_+}\leq C_0(x_0+C_1)$,  or  
$0<\rho^-(x_0)-\hat\rho_+\leq \frac{1}{-C_0(x_0+C_1)}$.  Thus $\rho^-(x_0)\rw\hat\rho_+$  when  $x_0\rw -\infty$.
Therefore  the difference  with nonextremal case $a<m$ is that 
  $\rho^-(x_0)-\hat\rho_+$  decays as  $O\big(\frac{1}{x_0}\big)$  in  the extremal case and $\rho^-(x_0)-\hat\rho_+$  decays
   exponentially in the  nonextremal  case.
   
   Consider the case when  $a^2>m^2$.   Then  the  outer and the inner  event horizons disappear  and this case is called  the case
   of naked singularity.  
Note that  the ergoregion  is $a<\rho<\sqrt{4m^2+a^2}$.  

\begin{theorem}																		\label{theo:9.1}
In the case  $a^2>m^2$,  i. e.  in the case  of the naked  singularity,  the null-geodesics  $\gamma^+$  starts  at a point  $P_0$     inside 
ergosphere  and  reaches 
 the singularity  ring  $\rho=a$  at some finite  time  $x_0^{(1)}>0$.  When  $x_0<0$  $\gamma^+$  tends  to the infinity  when  
$x_0\rw -\infty$.  
The null-geodesics   $\gamma^-$  also  starts  at  $P_0$  and  ends at the ring singularity at  a finite time 
$x_0^{(2)}>0$.
When  $x_0<0$  $\gamma^-$   first reaches  a turning point  $\rho_*$  at  $x^*<0$  when  $x_0$  decreases  and 
then it turns  and approaches  the singularity   at some finite time $x_0^{(3)}< x_0^*$  (see Fig. 2).
\end{theorem}
{\bf Proof:}
Consider  the behavior  of $\rho^-(x_0)$  when  $x_0<0$.  As in  (\ref{eq:7.26})  we can prove
  that there  exists  a turning point $\rho_*$  where
$a<\rho_0<\rho_*<\sqrt{4m^2+a^2}$  by showing that  $\Delta_2^-(\rho_0)>0$  and  $\Delta_2^-(\sqrt{4m^2+a^2}<0$. 
  Using that $\rho_0>a$,  we have   $\xi_0^-<\frac{2m-a}{2m+a}$.
Then $\frac{2m-a}{2m+a}<\frac{2a^2}{4m^2+a^2}$  when  $a>m$   and  this  proves that  $\Delta_2^-(\sqrt{4m^2+a^2})<0$ 
  (cf.  (\ref{eq:7.26})).
Therefore there exists  a turning point  $\rho_*$  where  $\Delta_2^-(\rho_*)=0$.  
Let  $x_0^{(3)}<0$  be  such   that  $\rho(x_0^{(3)}=\rho_*$.  When  $x_0<x_0^{(3)}$  the equation  for   $\rho^-(x_0)$
has the form 
$\frac{d\rho^-(x_0)}{dx_0}=\frac{\sqrt{\Delta_2^-}}{\sqrt{\Delta_1^-}}$  (cf.  (\ref{eq:7.27})).  Note that 
$1-Kb_\rho=\frac{(r-m)^2+a^2-m^2}{r^2+a^2}>\frac{a^2-m^2}{r^2+m^2}$.  Therefore  when  $a^2-m^2$  is  small and  $r$
  is close to $m$ we have
\begin{equation}																\label{eq:9.3}
\xi_\rho^-=O\Big(\frac{1}{a^2-m^2}\Big).                    
\end{equation}
Hence
\begin{equation}																\label{eq:9.4}
\frac{d\varphi}{d\rho}=O\Big(\frac{1}{a^2-m^2}\Big)
\end{equation}
is large  when  $a^2-m^2$  is small  and $r$  is close  to $m$  (cf.  (\ref{eq:7.29})).
Therefore $\gamma^-$    makes  several  rounds  near  $r=m$  and  the number  of these  rounds 
becomes larger  when  $a^2-m^2$
becomes smaller.  When  $x_0$  decreases  further
  $\gamma^-$  approaches the singularity  $\rho=a$.  Near $\rho=a$  we have,  using
(\ref{eq:8.8})  and  (\ref{eq:8.10}),  that  (cf.  (\ref{eq:8.11}))
\begin{equation}																\label{eq:9.5}
\frac{d\rho^-}{dx_0}=\frac{\sqrt{\Delta_2^-}}{\sqrt{\Delta_1^-}}\leq  \frac{C_0|\rho-a|^{-\frac{1}{4}}}{|\rho-a|^{-\frac{1}{2}}}
=C_0|\rho-a|^{\frac{1}{4}}.
\end{equation}
Similarly to  (\ref{eq:8.11})   we  have
$
\frac{4}{3}(\rho-a)^{\frac{3}{4}}\leq  C_0x_0+C.
$
Therefore 
\begin{equation}																\label{eq:9.6}
\rho^-(x_0)-a\leq  \frac{3}{4}C(x_0-x_0^{(4)})^{\frac{4}{3}},  \ \ x_0> x_0^{(4)},
\end{equation}
i.e.   $\rho(x_0^{(4)})=a$.  Also  $\big|\frac{d\varphi}{d\rho}\big|\leq C(\rho-a)^{-\frac{1}{4}}$.
Hence $|\varphi(x_0)-\varphi_0|\leq  C_0|\rho-\alpha|^{\frac{3}{4}}$.  Therefore  the  null-geodesics $\gamma^-$  ends at the point
$(a,\varphi_0)$  when  $x_0\rw x_0^{(2)}$  (see Fig.  2).

The behavior  of $\gamma^+$  and  $\gamma^-$  for  $x_0>0$   and  $\gamma^+$  for  $x_0<0$
is  the same  as  in  the case  $a^2<m^2$,  and the proof  is also the same.

\newpage

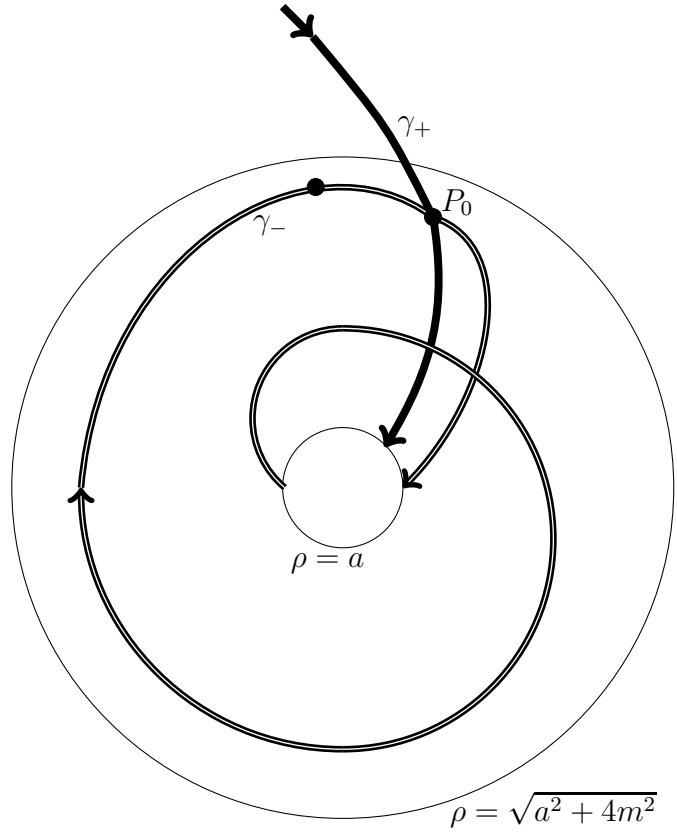
\begin{figure}
[ht]
\centering

\begin{tikzpicture}[scale=0.4]

\draw(0,0) circle  [radius = 2]; 
\draw(0,0) circle  [radius = 11];

\draw[<-][line width = 3 pt](1.4,1.35) .. controls   (3,4)      and (3.5,6) ..(3,9);
\draw[line width = 3 pt](3,9) .. controls (1.5,12)  ..(-1,15);
\draw[<-][line width = 3 pt] (-1,15) -- (-2,16);

\draw[<-][line width = 1 pt,double distance =0.5pt ](2,0) .. controls (5,3) and (6,8).. (3,9);
\draw[line width = 1 pt,double distance = 0.5pt](3,9) .. controls (-1,12) and (-8,8) ..  (-8.7,0);

\draw[<-][line width = 1 pt,double distance = 0.5pt](-8.7,0) arc (180:270:8.7);
\draw[line width = 1 pt,double distance = 0.5pt](0,-8.7) arc (270:450:7);
\draw[line width = 1 pt,double distance = 0.5pt](0,5.3) arc (90:230:3);

\filldraw[black] (-0.9,10 ) circle (8pt);

\filldraw[black] (3,9) circle (8pt);

\draw(-0.5,-2.5) node {$\rho=a$};
\draw(7,-10.7) node {$\rho=\sqrt{a^2+4m^2}$};

\draw(3.8,9.5)  node {$P_0$};
\draw(2.4,12) node {$\gamma_+$};
\draw(-2.4,8.7)  node  {$\gamma_-$};

\end{tikzpicture}
\caption{\label{myfigure-2} The case $a>m$.}
%
 Null-geodesics  $\gamma_+$  starts  at  $P_0$  and  reaches  $\rho=a$  at a finite  time  $x_0^{(1)}>0$.
When  $x_0\rw -\infty$  $\gamma_+$  tends  to   infinity.
 Null-geodesics   $\gamma_-$  starts  at $P_0$   and reaches  $\rho=a$  at  finite time $x_0^{(2)}>0$.
When  $x_0<0$   $\gamma_-$  passes a turning point, 
makes several rounds and reaches $\rho=a$  at some time  $x_0^{(3)}<0$.

 \end{figure}

\end{document}